\documentclass{document-template}
\usepackage{amsmath} 
\usepackage[T1]{fontenc}
\usepackage[utf8]{inputenc}
\usepackage[style=numeric, backend=bibtex, sorting=none]{biblatex}
\bibliography{references}
\usepackage{xcolor}
\usepackage{rotating}
\usepackage{tikz}
\usepackage{tcolorbox}
\usepackage{multirow}
\usepackage{minted}
\usepackage{listings}
\usepackage{fvextra} 
\usepackage{geometry}
\tcbuselibrary{listings}
\usepackage{listings}
\usepackage{pdfpages}
\usepackage{chngcntr}
\lstdefinestyle{method}{
  basicstyle=\ttfamily\small,
  frame=single,
  breaklines=true,
  columns=fullflexible,
  keepspaces=true,
  escapeinside={(*@}{@*)} 
}

\lstdefinestyle{inputbox}{
  basicstyle=\ttfamily\small,
  frame=single,
  breaklines=true,
  showstringspaces=false,
}

\lstdefinestyle{jsonbox}{
  basicstyle=\ttfamily\scriptsize,
  frame=single,
  breaklines=true,
  showstringspaces=false,
}

\tcbset{
  colback=white,
  colframe=black,
  sharp corners,
  boxrule=0.5pt
}

\begin{document}

\pagestyle{fancy}

\title{Ontology-aligned structuring and reuse of multimodal materials data and workflows towards automatic reproduction}

\maketitle

\author{Sepideh Baghaee Ravari}
\author{Abril Azocar Guzman}
\author{Sarath Menon}
\author{Stefan Sandfeld}
\author{Tilmann Hickel}
\author{Markus Stricker*}

\begin{affiliations}

S. B. R., S. M., M. S.\\
Interdisciplinary Centre for Advanced Materials Simulation, Ruhr-Universit\"at Bochum, Universit\"atsstra\ss e 150, 44801 Bochum, Germany\\
$^*$Email Address: markus.stricker@rub.de\\

A. A. G., S. S.\\
Institute for Advanced Simulations -- Materials Data Science and Informatics (IAS‑9), Forschungszentrum J\"{u}lich GmbH, 52425 J\"{u}lich, Germany\\

T. H.\\
Max Planck Institute for Sustainable Materials, 40237 D\"usseldorf, Germany\\
BAM Federal Institute for Materials Research and Testing, 12489 Berlin, Germany

\end{affiliations}

\keywords{text mining, workflow, large language models, stacking fault energy}

\begin{abstract}
Reproducibility of computational results remains a challenge in materials science, as simulation workflows and parameters are often reported only in unstructured text and tables.
While literature data are valuable for validation and reuse, the lack of machine-readable workflow descriptions prevents large-scale curation and systematic comparison.
Existing text-mining approaches are insufficient to extract complete computational workflows with their associated parameters.
An ontology-driven, large language model (LLM)-assisted framework is introduced for the automated extraction and structuring of computational workflows from the literature.
The approach focuses on density functional theory-based stacking fault energy (SFE) calculations in hexagonal close-packed magnesium and its binary alloys, and uses a multi-stage filtering strategy together with prompt-engineered LLM extraction applied to method sections and tables.
Extracted information is unified into a canonical schema and aligned with established materials ontologies (CMSO, ASMO, and PLDO), enabling the construction of a knowledge graph using atomRDF.
The resulting knowledge graph enables systematic comparison of reported SFE values and supports the structured reuse of computational protocols.
While full computational reproducibility is still constrained by missing or implicit metadata, the framework provides a foundation for organizing and contextualizing published results in a semantically interoperable form, thereby improving transparency and reusability of computational materials data.
\end{abstract}

\section{Introduction}

Trustworthy science largely relies upon the ability to reproduce reported results.
Although what exactly `reproducibility' and `trustworthiness' constitute in the scientific process is subject to discussion~\cite{Peels2023}.
In computational materials science, reproducibility is in principle solved through workflow management systems~\cite{Janssen2019,Huber2020} and efforts of the community are now increasingly concentrating towards homogenization of the \textit{language} to describe interoperable workflows~\cite{Janssen2025}.
However, these efforts largely relate to results which have been produced using a workflow management software in the first place and the condition that the corresponding workflows are published along with the results in an appropriate format.

Existing results in the scientific literature which have either been produced without workflow management software or without explicit publication of the implemented workflow are valuable sources of information for comparisons and validation of new results but hardly usable in their entirety because of the substantial manual effort involved in curating them into a coherent dataset that includes metadata. 
The latter cases share the information describing the ``workflow'' in the form of descriptive text and possibly tabular data.
(Semi-)automated extraction of such information from literature (text mining) has a long history, in which the field of materials science presents its own challenges~\cite{Kononova2021}.

Large Language Models (LLMs) are increasingly used to extract materials-related data from scientific texts~\cite{Schilling-Wilhelmi2025}, which potentially enables the creation of structured datasets from unstructured scientific articles.
To a large extent, existing use cases of LLMs either convert the latent knowledge of scientific texts to embeddings~\cite{Tshitoyan2019,Trewartha2022,Zhang2025a} and make use of the correlations in embedding space, or they concentrate on the extraction of individually reported material property values~\cite{Swain2016,Mavracic2021,Sipilae2024,Kelly2025} or they are very specific to a certain application~\cite{Katzer2025} and not easily generalizable.
One notable, very different example for semi-automated creation of a large and diverse dataset is ChemPile~\cite{Mirza2025b}, where source code and workflows were directly mined from Github. 
The intended use case of the latter approach, however, is not to reproduce any of the computations, but to improve the reasoning of LLMs during training or fine-tuning.
In short, none of the current methods allow the extraction of computational workflows including parameters based on their textual and tabular presentation in scientific articles.

This is what we define as our target: A methodology which automatically extracts necessary computational parameters from scientific texts to allow the (semi-) automated reproduction of a presented computational study.
\textit{Automatically} here is defined as `No other specific user input is required after the original creation of the extraction workflow'.
More specifically, here we target the extraction of computational workflows used for atomistic simulations based on density functional theory (DFT) of stacking fault energies (SFE) in Magnesium and its alloys. These defects are chosen because of their relevance to ductility~\cite{Wu2015} and the variety of simulation protocols to determine them.
To achieve this, we use a combination of Prompt Engineering of LLMs and ontology-alignment of the extracted, structured information with existing ontologies for the description of ``computational workflows''.
Subsequently, we use the structured information to create a Knowledge Graph.
This Knowledge Graph can be created using atomRDF~\cite{atomRDF}, a tool for ontology-based creation, manipulation, and querying of atomistic structures and their properties as well as their computational provenance.
With this approach we aim for the following three application:

\begin{itemize}
    \item The collection of existing results reported in journals for the sake of comparing them by querying the Knowledge Graph.
    \item The reproduction of published results to compare to own results using the Knowledge Graph in conjunction with atomRDF.
    \item The reuse of reported computational simulation protocols, e.g., for parameter studies to extend prior studies using the Knowledge Graph, atomRDF, and additional implementation.
\end{itemize}

The paper is organized as follows: In Section~\ref{sec:methods_data} we summarize the used ontologies, the basics of ontology-based workflows, the creation of our reference data set as well as the LLM-based extraction workflow. Section~\ref{sec:results} details how the concepts of existing ontologies need to be extended for our existing case. In addition, we present SPARQL queries of the knowledge graph. Section~\ref{sec:discussion} provides an assessment for the approach and we conclude in Section~\ref{sec:conclusion}.

\section{Methods and data}
\label{sec:methods_data}
\subsection{Ontologies}
Semantic web technologies provide a foundation for FAIR research data management by enabling structured, machine-readable descriptions of data and workflows. Beyond data management, ontologies and knowledge graphs support AI-ready datasets by expressing complex relationships, enabling structured querying, and facilitating downstream tasks such as reasoning and automated validation of reproducibility.

For the use case addressed in this work, we rely on established materials science ontologies to describe atomistic simulations and their workflows. The Computational Materials Sample Ontology (CMSO)~\cite{CMSO} and the Atomistic Simulation Methods Ontology (ASMO)~\cite{ASMO} provide classes and properties to represent the computational sample, modeling methods, and calculated properties at the atomistic level. Workflow execution and provenance are described using PROV-O~\cite{w3c-provo}, a W3C-recommended ontology for provenance representation. Furthermore, ASMO reuses terms from the Materials Design Ontology (MDO)~\cite{Li2020AnOF}. In addition, crystallographic defects are described using the Open Crystallographic Defect Ontologies (OCDO)~\cite{OCDO}, with stacking faults represented in the Planar Defect Ontology (PLDO)~\cite{PLDO}.

Because ontology coverage is an ongoing community effort, extensions at the domain and application level are required for this study. Accordingly, we introduce targeted ontology extensions to capture the domain-specific concepts required for stacking fault simulations and their automated extraction and integration within ontology-aligned workflows.

\subsection{Semantic annotation in workflows}
A simulation workflow can be defined as an abstract layer on top of individual software components~\cite{bekemeier2025advancing}.
I.e. an ordered sequence of computational steps with specific input data, simulation methods, and post-processing operations that are combined to produce a well-defined set of outputs.
Each step in the workflow is explicitly documented, including its required inputs, applied computational procedures, and generated results. By formalizing simulation processes in this manner, workflows enable transparency, reproducibility, and reuse of computational studies, while reducing the need for detailed knowledge of individual simulation tools or implementation details~\cite{10.1007/978-3-030-62466-8_14,Wilkinson2025}, this is achieved by the use of a workflow management system (WfMS).

Note, `a workflow' can also be a textual description of a simulation protocol. 
A workflow is not necessarily semantically aligned with all other workflows.
A common language for workflows, i.e. a semantic annotation, however, provides a path to seamless reproducibility.

Workflow management systems provide the software infrastructure required to define, execute, and manage such simulation workflows. In materials science, platforms such as Pyiron or AiiDA~\cite{janssen2019pyiron,Huber2020,Huber2022,Janssen2025} offer an integrated environment for constructing modular and reusable workflows, handling simulation data, tracking provenance, and orchestrating complex multi-step calculations. By abstracting low-level execution details and providing standardized interfaces to atomistic simulation codes, workflow managers like Pyiron facilitate scalable, reproducible, and data-driven simulation studies.

We ensure semantic interoperability by combining a workflow manager with the atomRDF software framework~\cite{atomRDF}, it connects computational workflows with semantic data integration by transforming workflow-generated metadata into structured knowledge graphs.
An alternative solution to ensure semantic interoperability could be the use of \texttt{semantikon}~\cite{Dareska2025} to provide scientific context to Python functions.
During workflow execution, relevant metadata are captured in lightweight conceptual dictionaries, which atomRDF converts into RDF triples aligned with application ontologies, specifically CMSO~\cite{CMSO} and ASMO~\cite{ASMO}, together with PROV-O~\cite{w3c-provo} for provenance. This ensures a consistent and machine-readable representation of computational samples, simulation methods, and workflow execution steps. By embedding semantic annotation at the workflow level, atomRDF enables reproducible and interoperable data generation while remaining agnostic to the underlying workflow engine.

\subsection{Ontology-based SFE workflows}\label{ssec:onto_sfe}
While reproducible workflows ensure that computational results can be regenerated, semantic data annotation is required to ensure that these results are interpretable, reusable, and interoperable beyond the context of the original workflow. 
Because the existence of a workflow, implicitly given by the description in an article or an explicit implementation in code, does not necessarily mean that `a system' (computer or human) can understand what the workflow does because there is a huge freedom, e.g. in variable and function naming.
In this work, both aspects are addressed by integrating ontology-aligned metadata generation directly at the source, namely during workflow execution.

\subsubsection{SFE workflows}\label{sssec:sfe_workflows}
First, we present three simulation workflows within the Pyiron framework~\cite{janssen2019pyiron} to calculate SFE using the rigid shift method, the tilted-cell method, and the axial next-nearest neighbor Ising (ANNNI) analytical model. 

In the following, we briefly describe the theoretical background and practical implementation of each of these SFE calculation approaches.

\begin{itemize}
    \item Rigid shift method\\
    In the rigid shift method, a stacking fault is created by rigidly displacing one half of a supercell relative to the other along a specified slip plane and direction. A supercell is constructed with a fault plane, and a displacement vector $\mathbf{u}$ is applied parallel to this plane.
    The SFE is defined as
    \begin{equation}
    \gamma(\mathbf{u}) = \frac{E(\mathbf{u}) - E_0}{A},
    \end{equation}
    where $E(\mathbf{u})$ is the total energy of the displaced configuration, $E_0$ is the energy of the perfect crystal, and $A$ is the area of the fault plane. By sampling $\mathbf{u}$ over the primitive translation vectors of the plane, a full $\gamma$-surface is obtained~\cite{Vitek1968}.
    \item Tilted-cell method\\
    In the tilted-cell method, stacking faults are introduced through a homogeneous deformation of the simulation cell rather than an explicit rigid displacement of atoms. The lattice vectors are tilted such that periodic boundary conditions enforce a stacking mismatch across the fault plane.
    After applying the appropriate deformation, atomic positions are relaxed while maintaining the modified cell geometry. The SFE is then obtained from the excess energy per unit area relative to the perfect crystal similar to rigid shift method~\cite{yin2017comprehensive}. 
    \item ANNNI model\\
    Within the framework of the ANNNI model, the total energy of a crystal composed of a sequence of stacked atomic planes is expressed as a sum of interplanar interaction energies, together with a reference term corresponding to the energy of non-interacting planes:
    \begin{equation}
    E = E_0 - \sum_{n=1}^{N} J_n \sum_{i=1}^{M} s_i s_{i+n},
    \end{equation}
    where $s_i = \pm 1$ denotes the stacking state of the $i$-th atomic plane, $J_n$ are the interplanar interaction parameters, and $M$ is the total number of planes.
    
    As an example, the energies per atomic plane of the ideal hexagonal close-packed (HCP), face-centered cubic (FCC), and double hexagonal close-packed (DHCP) stacking sequences can be written as
    
    \begin{align}
    E_{\mathrm{HCP}}  &= J_{0} - J_{1} - J_{2} - O(J_{3}), \\
    E_{\mathrm{FCC}}  &= J_{0} + J_{1} - J_{2} + O(J_{3}), \\
    E_{\mathrm{DHCP}} &= J_{0} + J_{2} + O(J_{3}),
    \end{align}
    
    where $J_{0} = E_{0}/M$ denotes the energy of a single atomic plane in the absence of interplanar interactions.
    
    Using these expressions, the intrinsic stacking fault energies of HCP structure are obtained by comparing the energies of the corresponding ideal stacking sequences and normalizing by the area of a single HCP basal plane, $A = \frac{\sqrt{3}}{2}a^2$:
    \begin{align}
    \gamma_{\mathrm{I1}} &= \frac{2\left(E_{\mathrm{DHCP}} - E_{\mathrm{HCP}}\right)}{A}, \label{eq:eq11} \\
    \gamma_{\mathrm{I2}} &= \frac{E_{\mathrm{FCC}} + 2E_{\mathrm{DHCP}} - 3E_{\mathrm{HCP}}}{A}. \label{eq:eq12}
    \end{align}

    Accordingly, stacking fault energies are determined entirely from the energies of perfect crystal structures such as HCP, FCC, and DHCP, without explicitly introducing stacking faults into the simulation cell~\cite{ruffino2020ising}.

\end{itemize}

\subsubsection{Ontology extensions}\label{sec:onto_ext}
The semantic description of the computational aspects of simulations is to a large extent covered by the CMSO~\cite{CMSO} and ASMO~\cite{ASMO} ontologies.
However, for the present use case of stacking faults in magnesium, extensions are required, mainly to represent HCP systems.
These extensions include commonly used labels of crystallographic planes such as basal, pyramidal and prismatic planes, which are necessary for an unambiguous description of stacking fault geometries.
We extend the ASMO~\cite{ASMO} ontology to capture additional modeling details relevant to the simulations.
This includes a representation of the ANNNI model as a simulation algorithm and an expanded set of structural operations that can be applied to supercells during atomistic workflows.
These operations include rotation, translation, and shear, all defined as subclasses of a generic spatial transformation.
Most extensions focus on the semantic description of stacking faults, which is provided by the PLDO~\cite{PLDO} ontology, a module of the OCDO~\cite{OCDO} ontology.
New classes are introduced to categorize different types of stacking faults, including intrinsic, extrinsic, twin-like, and specific I1 and I2 stacking faults, as well as classes representing the stacking fault energy and the generalized stacking fault energy.
In addition, object and data properties ware added to describe stacking sequences, fault planes, and displacement vectors.
Finally, the ontology is extended to represent defect complexes, enabling the description of interactions between solute atoms and stacking faults.
This includes properties to capture the relative distance between solutes and defect planes, as well as the local solute concentration.

\begin{figure}[H]
    \centering
    \includegraphics[width=\linewidth]{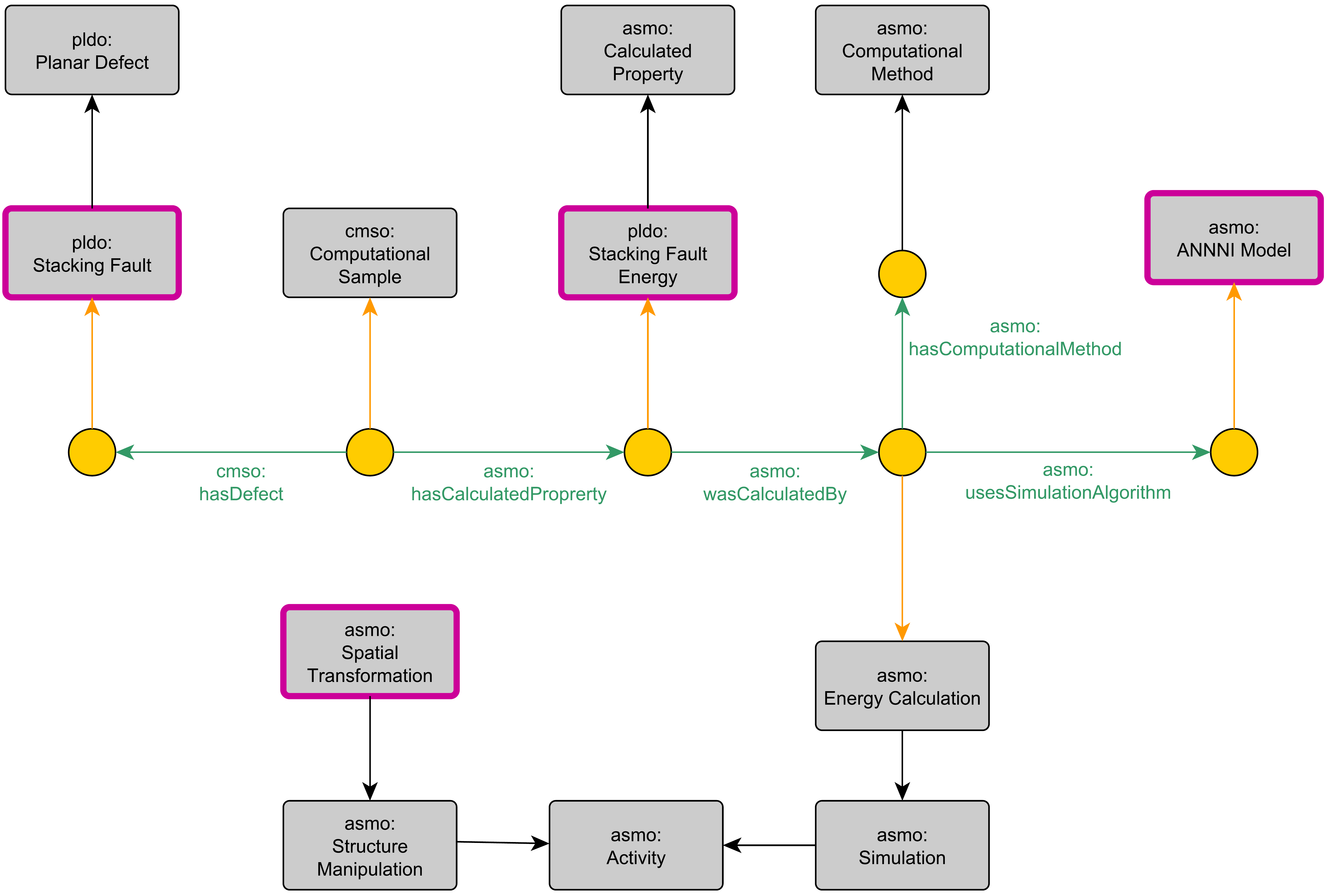}
    \caption{Overview of ontological SFE workflow description.}
    \label{fig:SFE_workflow_onto}
\end{figure}

A high-level overview of the ontology-based SFE workflows is shown in Figure~\ref{fig:SFE_workflow_onto}. 
For the ANNNI model workflow, the process consists of an energy calculation simulation activity, which uses a specific simulation algorithm.
In contrast, the tilted-cell and rigid-shift approaches involve an energy calculation simulation activity that is executed in parallel with an operation on the computational sample's simulation cell. This operation corresponds to one of the three defined spatial transformations.
In both cases, each simulation is associated with a computational method and produces an output in the form of the SFE. 
The resulting SFE is also linked to a computational sample, which is characterized by a specific type of stacking fault, as described above.
This representation are then stored in atomRDF within Pyiron framework. 

However a more detailed instantiation of this abstract representation is given in the following JSON example, which explicitly encodes the entities and relations involved in a SFE calculation with  the rigid shift method.
The JSON represents the minimum set of information required to describe an SFE workflow within the proposed ontological framework, including the computational samples, defect description, calculated property, and associated computational method.
While sufficient to capture the essential structure of the workflow, this representation is intentionally simplified for the ontology-alignment step from the literature; the full atomRDF JSON contains a more detailed and fine-grained description of materials, processes, and parameters.
The differences between this minimal representation and the complete atomRDF schema are discussed in detail in the Discussion section.

\vspace{1em}

\noindent
\begin{minipage}[t]{0.40\textwidth}
\textbf{Ontology-based SFE workflow JSON representation:}
{\renewcommand{\baselinestretch}{0.60}
\begin{lstlisting}[style=jsonbox]
{
 "ComputationalSample_1": {
    "ChemicalSpecies": {
      "ChemicalElement": ,
      "SubstitutionalSolute": ,
      "hasSoluteConcentration": 
    },
    "ChemicalComposition": ,
    "Material": ,
    "CrystalStructure": ,
    "SimulationCell": {
      "SimulationCellVector": {
        "hasComponent_x": ,
        "hasComponent_y": ,
        "hasComponent_z":
      },
      "SimulationCellLength": {
        "hasLength_x": ,
        "hasLength_y": ,
        "hasLength_z": 
      },
      "SimulationCellAngle": {
        "hasAngle_alpha": ,
        "hasAngle_beta": ,
        "hasAngle_gamma": 
      },
      "hasRepetition_x": ,
      "hasRepetition_y": ,
      "hasRepetition_z": ,
      "hasUnitCellType": ,
      "hasVacuumLength": ,
      "hasVacuumDirection": 
    }},
  "ComputationalSample_2": {
    "ChemicalSpecies": {
      "ChemicalElement": ,
      "SubstitutionalSolute": ,
      "hasSoluteConcentration": 
    },
    "ChemicalComposition": ,
    "Material": ,
    "CrystalStructure": ,
    "SimulationCell": {
      "SimulationCellVector": {
        "hasComponent_x": ,
        "hasComponent_y": ,
        "hasComponent_z": 
      },
      "SimulationCellLength": {
        "hasLength_x": ,
        "hasLength_y": ,
        "hasLength_z": 
      },
      "SimulationCellAngle": {
        "hasAngle_alpha": ,
        "hasAngle_beta": ,
        "hasAngle_gamma": 
      },
      "hasRepetition_x": ,
      "hasRepetition_y": ,
      "hasRepetition_z": ,
      "hasUnitCellType": ,
      "hasVacuumLength": ,
      "hasVacuumDirection": 
    },
    "StackingFault": {
      "Type": ,
      "hasDisplacementVector": ,
      "hasSFplane": ,
      "hasStackingSequence": ,
      "SF_hcp_label": 
    }},
  "I1StackingFaultEnergy: {
    "value": 
    "unit": "
    "label": 
  },
  "ComputationalMethod": {
    "Simulation": ,
    "Type": ,
    "SoftwareAgent": ,
    "XC_EnergyFunctional":,
    "EnergyCutoff": {
      "value":,
      "unit": "
    },
    "KpointMesh": {
      "Type": ,
      "value":
    },
    "InputParameter": {
      "Pseudopotential": {
        "label": 
      },
      "EnergyConvergence": {
        "value": 
        "unit": ,
        "label":
      },
      "ForceConvergence": {
        "value": 
        "unit": 
        "label": 
      }
    }
  }
}
    

\end{lstlisting}}
\end{minipage}

\subsection{Literature extraction workflow}
A high-level overview of our workflow is provided here. Further details are provided along with the results because we believe that the presentation of our methodological choices benefits from examples.
In this work, all language model–based processing steps are performed using GPT-4.1~\cite{openai2024gpt4technicalreport} as the underlying large language model to which we will refer to as LLM in the following.
For tasks requiring different levels of determinism or semantic variability, we adjust the sampling temperature parameter, hereafter denoted by the symbol~$T$.
Throughout the remainder of this paper, we explicitly indicate the value of $T$ used for each LLM call.

\begin{figure}[H]
    \centering
    \includegraphics[width=0.80\linewidth]{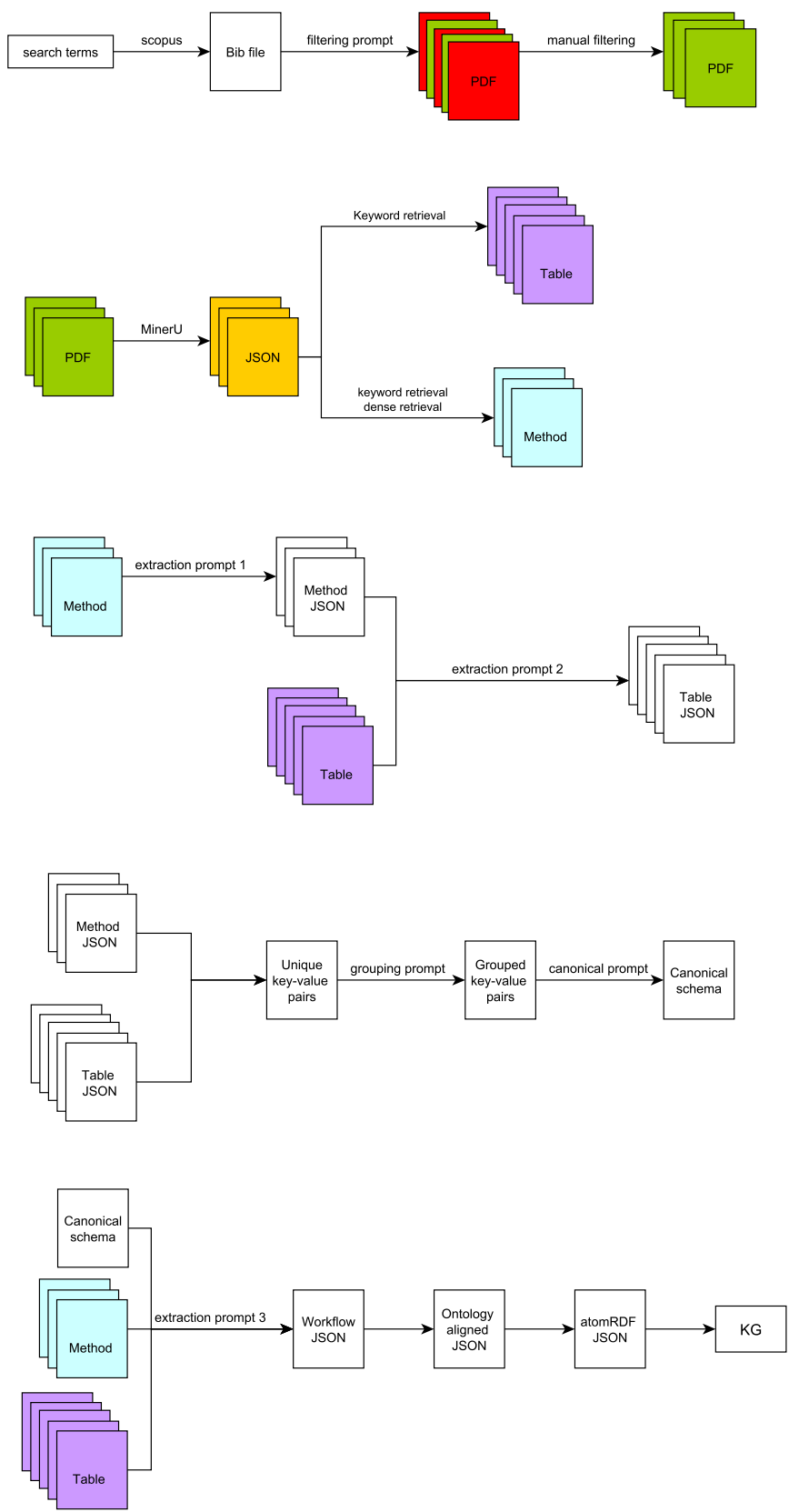}
    \caption{Schematic of LLM-based data extraction and normalization workflow of SFE calculations.}
    \label{fig:workflow}
\end{figure}

As shown in Figure~\ref{fig:workflow}, we first construct a relevant corpus through a Scopus search focused on SFE. This corpus is then filtered using a classification prompt (Appendix~\ref{lst:filtering}) applied to the titles and abstracts, employing LLM ($T=0$). Only papers reporting computational studies based on Density Functional Theory (DFT) for pure Mg and binary Mg-based alloys with a HCP structure, and explicitly addressing parameters influencing stacking faults, are retained. The selected articles are subsequently downloaded.

Next, a manual filtering step is performed to organize and identify the parameters that influence SFE. The articles, available in PDF format, are then parsed using the MinerU~\cite{wang2024mineruopensourcesolutionprecise} tool. For each paper, the methods section and the tables specifically related to SFE are retrieved based on keywords or optimized dense retrieval approach.

Using Extraction Prompt~1 (Appendix~\ref{lst:Extraction_prompt_1}) and LLM ($T=0$), we extract the computational workflow for SFE calculations from the method section of individual paper and represent it in JSON format. For each relevant table, we then use the corresponding keys from the method-section JSON together with Extraction Prompt~2 (Appendix~\ref{lst:Extraction_prompt_2}) and LLM ($T=0$) to extract stacking fault energies and associated parameters, producing enriched JSON representations of tables.

Once the JSON files from both method sections and tables are obtained, we collect all unique key-value pairs and use LLM ($T=0.3$) to group them semantically. Subsequently, the LLM ($T=0.5$) is used to define the minimum set of canonical keys capable of representing each semantic group. This process yields a generalized template for atomistic stacking fault energy computation workflows of stacking fault energy by unifying the keys across all JSON representations.

The canonical schema is then manually aligned with ontology classes and object properties from the CMSO~\cite{CMSO}, ASMO~\cite{ASMO}, and PLDO~\cite{PLDO} ontologies. However, for data extraction across all papers, the canonical schema rather than the ontology-aligned schema is employed. Accordingly, for each combination of method section and corresponding tables in a paper, we apply the canonical schema using Extraction Prompt~3 (Appendix~\ref{lst:Extraction_prompt_3}) and LLM ($T=0.3$).

Finally, ontology-aligned JSON files are converted into a knowledge graph by extracting triples using the atomRDF package~\cite{atomRDF}.

\subsection{Reference dataset}
We create two manually curated reference datasets. The first dataset is used for the manual classification of papers according to the automatic filtering criteria. The second dataset contains individual data points extracted from tables, each linked to the corresponding methodological information from the same paper using an ontology-aligned JSON representation.

\section{Results}
\label{sec:results}
\subsection{Dataset creation}
\subsubsection{Scopus search}
Using SerialSearch option of Pybliometrics package~\cite{ROSE2019100263}, we conduct a comprehensive Scopus search (20\textsuperscript{th} July 2025) restricted to `Research articles' written in English language.
The search is performed on the title, abstract, and keywords fields using three layers of criteria: 
\begin{enumerate}
    \item Defect search terms:  
    \begin{quote}
    [`` stacking fault* '' OR `` planar fault*'' OR `` stacking-fault* ''  OR `` planar-fault*'']  
    \end{quote}
    This results in 21,080 papers.
    
    \item Defect property search terms:  
    \begin{quote}
    [ ``stacking fault* energ*'' OR ``fault *energ*'' OR ``planar fault* energ*'' OR `` stacking-fault* energ*'' OR `` planar-fault* energ*'']  
    \end{quote}
    These criteria reduce the number of papers to 5,607.

    \item Material search terms:  
    \begin{quote}
    [``Mg'' OR ``magnesium'' OR ``Magnesium'' ]  
    \end{quote}
    This final filter further reduces the corpus to 379 papers, which serve as the initial dataset.
\end{enumerate}

\subsubsection{Paper filtering}
\begin{figure}[H]
    \centering
    \includegraphics[width=\linewidth]{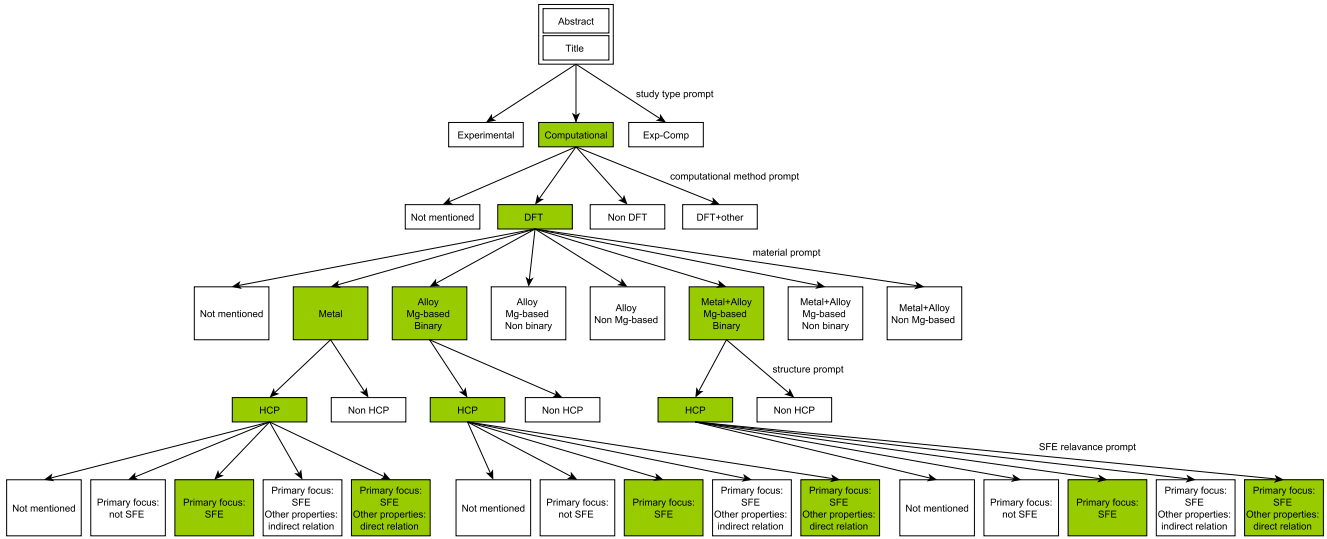}
    \caption{Decision-making process for automatic filtering.}
    \label{fig:automated_filtering}
\end{figure}

Starting from the 200 papers most relevant to our search criteria for the initial corpus, we employ another two-stage filtering strategy using papers' title and abstract. This filtering consists of a combination of automated screening and manual reviewing.
The objective of this procedure is to construct a dataset in which the dependencies between computational parameter choices for SFE calculations are similar across papers. 
Unlike most approaches to structured information extraction, which begin with a predefined JSON schema or Pydantic model~\cite{Pydantic2025}, our aim is to derive this schema directly from the literature. 
This requires identifying papers that not only compute SFE but do so in a comparable and interpretable way, in other words, we aim for a narrow methodological scope.
As shown in Figure~\ref{fig:automated_filtering}, we employ a hierarchical, multi-prompt classification pipeline using LLM ($T=0$) to automatically identify papers satisfying the following conditions:

\begin{itemize}
    \item Study type: Computational 
    \item Computational method: DFT 
    \item Material: Pure metals or binary Mg-based alloys 
    \item Structure: HCP
    \item Relevance to SFE: SFE is the major calculated property or/and other calculated properties have direct relation to SFE. 
\end{itemize}

\begin{figure}[htbp]
  \centering
  \includegraphics[width=\linewidth]{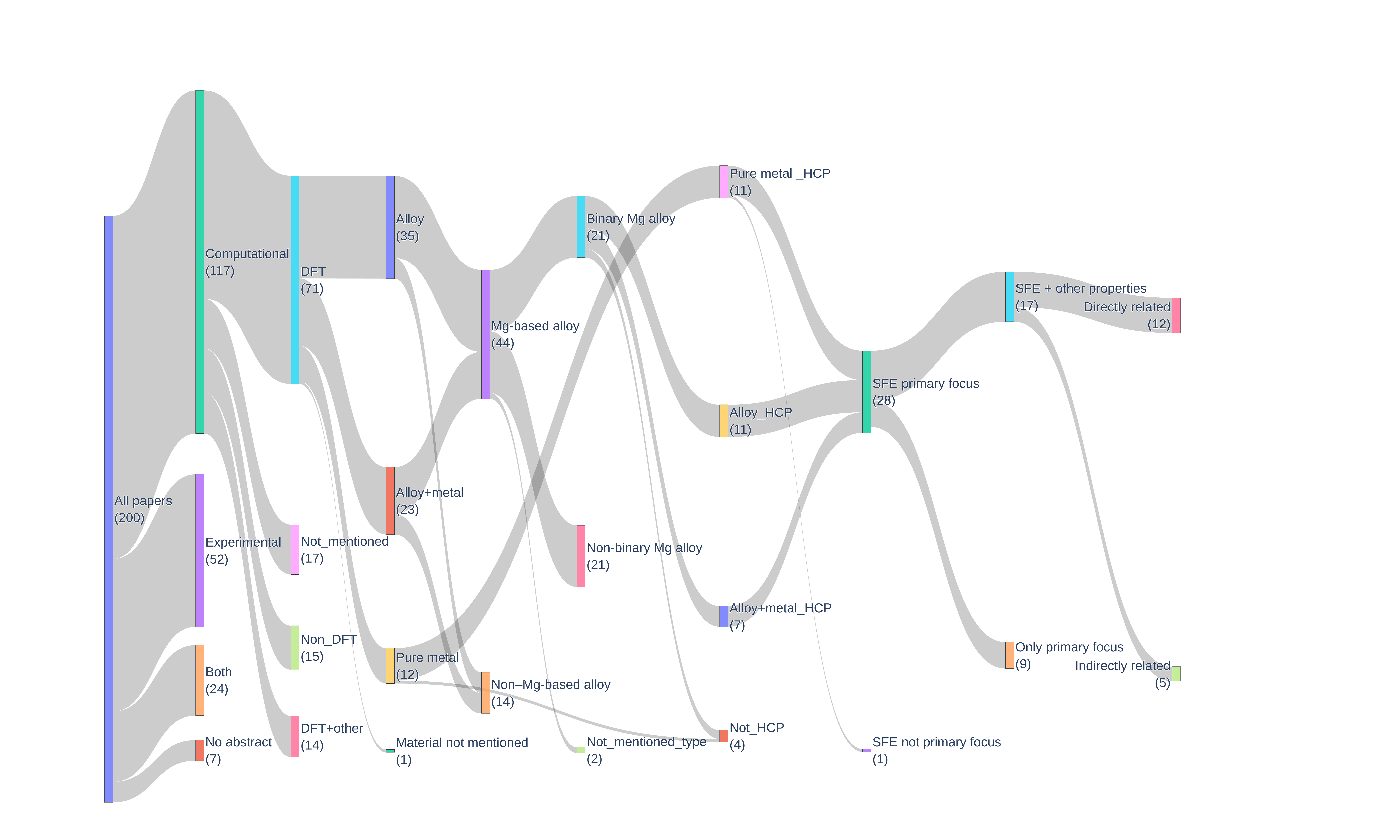}
  \caption{Hierarchical filtering and classification of the corpus in form of a Sankey diagram. 
  The Sankey diagram illustrates the sequential application of abstract availability, 
  study type, computational method, material system, compositional complexity, crystal structure, 
  and SFE relevance criteria, with node widths proportional to the number 
  of studies passing each stage. The height of each gray bar w.r.t. the left most purple bar corresponds to the fraction of paper retained after each filter.}
  \label{fig:sankey_pipeline}
\end{figure}

As shown in Figure~\ref{fig:sankey_pipeline}, the classification pipeline is a sequence of hierarchical decisions, each implemented via a dedicated prompt. Starting from an initial corpus of 200 papers, abstract availability is first assessed. Seven papers (3.5\%) lack an abstract and are excluded, leaving 193 papers (96.5\%) for further classification. These papers are then categorized by study type, yielding 117 computational studies (58.5\%), 52 experimental studies (26.0\%), and 24 studies (12.0\%) combining computational and experimental approaches. Only the computational studies are retained for subsequent filtering steps.

Within the computational subset, the primary computational methodology is examined. Of the 117 computational papers, 71 studies (60.7\%) employ density functional theory (DFT), while 14 studies (12.0\%) combine DFT with other approaches, 15 studies (12.8\%) rely exclusively on non-DFT methods, and 17 studies (14.5\%) do not explicitly specify the computational method. Only the 71 DFT-based studies are retained for material-level analysis.

Among these DFT studies, the investigated material systems comprise 35 alloy-only studies (49.3\%), 23 studies involving both alloys and pure metals (32.4\%), 12 studies focusing exclusively on pure metals (16.9\%), and one study (1.4\%) in which the material type is not specified. For alloy-containing studies, Mg-based systems are identified: a total of 44 studies involved Mg-based alloys, while 14 alloy studies are classified as non-Mg-based and excluded at this stage.

The Mg-based alloy studies are further subdivided according to compositional complexity, yielding 21 studies investigating binary Mg-based alloys, 21 studies examining non-binary systems, and two studies for which the alloy type is not specified. Only the binary Mg-based alloy systems are retained for further document-structural screening. Examination of crystal structure reveals that 11 pure-metal studies, 11 Mg-based binary alloy studies, and seven alloy-plus-metal Mg-based binary studies adopt a HCP crystal structure, while four studies are identified as non-HCP and excluded.

Following all compositional and structural filters, the final retained dataset consists of 29 HCP-based studies. Among these, 28 papers explicitly treat the SFE as a primary focus, while one studies address related properties without treating SFE as the central quantity of interest. The SFE-focused studies are further subdivided based on whether SFE constitutes the sole major calculated property (nine papers) or if it is studied alongside additional properties (17 papers). In the latter group, 12 studies examine properties directly related to SFE, while five studies consider properties indirectly related through additional modeling or analysis frameworks.

Combining the nine papers in which SFE constitute the sole major calculated property with the 12 papers in which SFE was studied alongside other properties that are directly related, a total of 21 papers are retained for further in-depth analysis. These studies represent cases where SFE plays a central and explicitly interpreted role in the reported results.

A detailed manual screening of these papers is then conducted to reduce conceptual complexity, harmonize terminology across different studies, and identify the core factors governing stacking fault energetics. Through this refinement process, the dataset is further narrowed to eight representative papers that provide clear, comparable, and methodologically transparent insights into SFE behavior.
The selected works focus on the following key aspects:

\begin{itemize}
    \item The influence of different slip planes and stacking fault types in HCP materials
    \item Variations in stacking fault modeling approaches, including the ANNNI model, rigid-shift methods, and tilted-cell techniques
    \item The role of structural relaxation procedures in determining fault energies
    \item The effect of individual alloying atoms positioned in the vicinity of the stacking fault plane
\end{itemize}

Table~\ref{tab:confusion_by_category} summarizes per-category confusion statistics (one-vs-rest) obtained by comparing the pipeline predictions against the reference annotations at each hierarchical stage. Here, the evaluation set size $N$ decreases along the pipeline because each subsequent decision is only applied to the subset of papers retained by the preceding filters. For each category, \textit{Reference} indicates how frequently the category occurs in the reference set within that step, while TP, TN, FP, and FN quantify correct positives, correct negatives, false positives, and false negatives, respectively.
At the study-type level ($N$=200), the pipeline shows strong agreement for computational ($\mathrm{TP} = 113$, $\mathrm{FN} = 1$) and both ($\mathrm{TP} = 20$, $\mathrm{FN} = 0$), while most discrepancies occur for experimental studies ($\mathrm{TP} = 52$, $\mathrm{FN} = 7$). Within the computational subset ($N$=113), the main source of error lies in separating DFT from adjacent categories: DFT exhibits $\mathrm{TP} = 62$ with $\mathrm{FP} = 9$ and $\mathrm{FN} = 5$, and DFT+other shows comparable ambiguity ($\mathrm{TP} = 9$, $\mathrm{FP} = 5$, $\mathrm{FN} = 5$), reflecting challenges in mixed-method reporting.

Material classification within DFT studies ($N$=62) is highly consistent across alloy ($\mathrm{TP} = 31$), alloy+metal ($\mathrm{TP} = 18$), and pure\_metal ($\mathrm{TP} = 10$), with only isolated misclassifications. Subsequent filters show similarly robust performance for Mg-based identification ($N$=49), alloy-system typing ($N$=37), and HCP structure screening ($N$=19). Downstream SFE-related labels achieve perfect agreement for SFE primary focus ($N$=16), whereas the final distinction between directly and indirectly related additional properties ($N$=8) remains the most ambiguous step ($\mathrm{FP} = 2$ and $\mathrm{FN} = 2$), reflecting conceptual overlap in multi-property studies.

\begin{table}[t]
\centering
\caption{Per-category confusion statistics (one-vs-rest) comparing pipeline predictions against reference annotations. $N$ denotes the number of papers evaluated at each hierarchical step after filtering by preceding decisions.}
\label{tab:confusion_by_category}
\small
\setlength{\tabcolsep}{5pt}
\begin{tabular}{llrrrrrr}
\hline
Step & Category & $N$ & Reference & TP & TN & FP & FN \\
\hline

\multirow{4}{*}{\parbox{4.2cm}{\textit{Step 1:}\\Study type}}
& both           & 200 & 20  & 20  & 176 & 4 & 0 \\
& computational  & 200 & 114 & 113 & 82  & 4 & 1 \\
& experimental   & 200 & 59  & 52  & 141 & 0 & 7 \\
& No abstract    & 200 & 7   & 7   & 193 & 0 & 0 \\
\hline

\multirow{4}{*}{\parbox{4.2cm}{\textit{Step 2:}\\Computational method}}
& DFT            & 113 & 67 & 62 & 37 & 9 & 5 \\
& DFT+other      & 113 & 14 & 9  & 94 & 5 & 5 \\
& Non\_DFT       & 113 & 16 & 15 & 97 & 0 & 1 \\
& Not\_mentioned & 113 & 16 & 10 & 94 & 3 & 6 \\
\hline

\multirow{4}{*}{\parbox{4.2cm}{\textit{Step 3:}\\Material type}}
& pure\_metal    & 62 & 11 & 10 & 51 & 0 & 1 \\
& alloy          & 62 & 32 & 31 & 29 & 1 & 1 \\
& alloy+metal    & 62 & 18 & 18 & 43 & 1 & 0 \\
& Not\_mentioned & 62 & 1  & 1  & 61 & 0 & 0 \\
\hline

\multirow{3}{*}{\parbox{4.2cm}{\textit{Step 4:}\\Mg-based alloy}}
& Yes            & 49 & 37 & 37 & 11 & 1 & 0 \\
& No             & 49 & 12 & 11 & 37 & 0 & 1 \\
& & & & & & & \\
\hline

\multirow{3}{*}{\parbox{4.2cm}{\textit{Step 5:}\\Alloy system type}}
& binary         & 37 & 21 & 19 & 16 & 0 & 2 \\
& non\_binary    & 37 & 16 & 16 & 20 & 1 & 0 \\
& Not\_mentioned & 37 & 0  & 0  & 36 & 1 & 0 \\
\hline

\multirow{3}{*}{\parbox{4.2cm}{\textit{Step 6:}\\Crystal structure}}
& HCP            & 19 & 16 & 16 & 3  & 0 & 0 \\
& Not\_HCP       & 19 & 3  & 3  & 16 & 0 & 0 \\
& & & & & & & \\
\hline

\multirow{3}{*}{\parbox{4.2cm}{\textit{Step 7:}\\SFE primary focus}}
& Yes            & 16 & 16 & 16 & 0  & 0 & 0 \\
& No             & 16 & 0  & 0  & 16 & 0 & 0 \\
& & & & & & & \\
\hline

\multirow{3}{*}{\parbox{4.2cm}{\textit{Step 8:}\\SFE only primary focus}}
& No             & 16 & 8 & 8 & 8 & 0 & 0 \\
& Yes            & 16 & 8 & 8 & 8 & 0 & 0 \\
& & & & & & & \\
\hline

\multirow{3}{*}{\parbox{4.2cm}{\textit{Step 9:}\\Other properties}}
& Directly\_related   & 8 & 3 & 3 & 3 & 2 & 0 \\
& Indirectly\_related & 8 & 5 & 3 & 3 & 0 & 2 \\
& & & & & & & \\
\hline

\end{tabular}
\end{table}

\subsubsection{PDF parsing}
Extracting information directly from PDF files is exceptionally difficult because PDF is a layout-oriented format rather than a structured text format.
PDFs store characters with precise positions on the page but do not encode semantic elements such as words, sentences, paragraphs, section headings, or table structure. As a result, text extraction tools must reconstruct logical structure from low-level layout cues, which often leads to errors such as broken reading order, mixed columns, misplaced captions, or fragmented tables. To enable reliable downstream processing such as LLM-based parsing or structured extraction, we convert PDFs into a more semantically meaningful format such as Markdown and JSON, where document hierarchy and text flow are explicitly represented~\cite{Bast2017BenchmarkPDFExtraction}.
This conversion serves two purposes: (1) to identify the document layout and its different content modalities, and (2) to transform the PDF text into a markup language. 
In this work, we use MinerU~\cite{wang2024mineruopensourcesolutionprecise} to identify and extract key modalities from scientific PDFs, including text, figures, tables, and equations, and to convert them into structured markup.
MinerU~\cite{wang2024mineruopensourcesolutionprecise} produces a Markdown file along with multiple JSON files. Among these, the `content list JSON' provides a structured representation of the document elements, where each entry corresponds to a specific modality (e.g., text, image, table, or equation) and includes associated metadata such as page indices, captions, footnote and extracted content.
For the purposes of our workflow, however, we rely exclusively on two modalities: text and tables, which contain the information relevant to downstream extraction and analysis.

\subsection{Information retrieval}
The objective of the information retrieval pipeline is to retrieve both (i) the computational workflow description used to calculate SFE and (ii) the numerical SFE values reported in the paper. Workflow descriptions typically appear in a Method section, whereas SFE values may appear in the Results section in the form of text, figures, or tables. Here, we focus on retrieving the Method section describing the SFE computation workflow and tables containing SFE values, which serve as structured data sources.
We employ two strategies detailed below.

\subsubsection{Keyword-based retrieval}

\paragraph{Method section}\leavevmode\par
For papers that follow the structure of a typical scientific article of Introduction-Methods-Results-Discussion (IMRD), we identify the method section by scanning ``type'' key of the content list JSON for the root terms:
\textit{``Method''}, \textit{``Computation''}, \textit{``Simulation''}, and \textit{``Calculation''}.
The first occurrence of any of these terms marks the beginning of the Method section.
To determine the end of the section, we locate the subsequent header containing the root term \textit{``Result''}.
We then define the text between these headers as the `method description'.

\paragraph{Tables}\leavevmode\par
For the tables, we first identify all table structures detected in the PDF using the content list JSON. We filter objects where the field ``type'' is set to ``table'', and for each table we extract its body, caption, and any associated footnotes. We then retain only those tables whose captions contain SFE-related keywords, including \textit{``stacking fault energy''}, \textit{``SFE''}, \textit{``GPFE''}, \textit{``generalized planar fault energy''}, \textit{``GSFE''}, and \textit{``generalized stacking fault energy''}. Only tables matching at least one of these keywords are kept for further analysis.

\subsubsection{Dense retrieval}
\begin{figure}[htbp]
  \centering
  \includegraphics[width=0.70\linewidth]{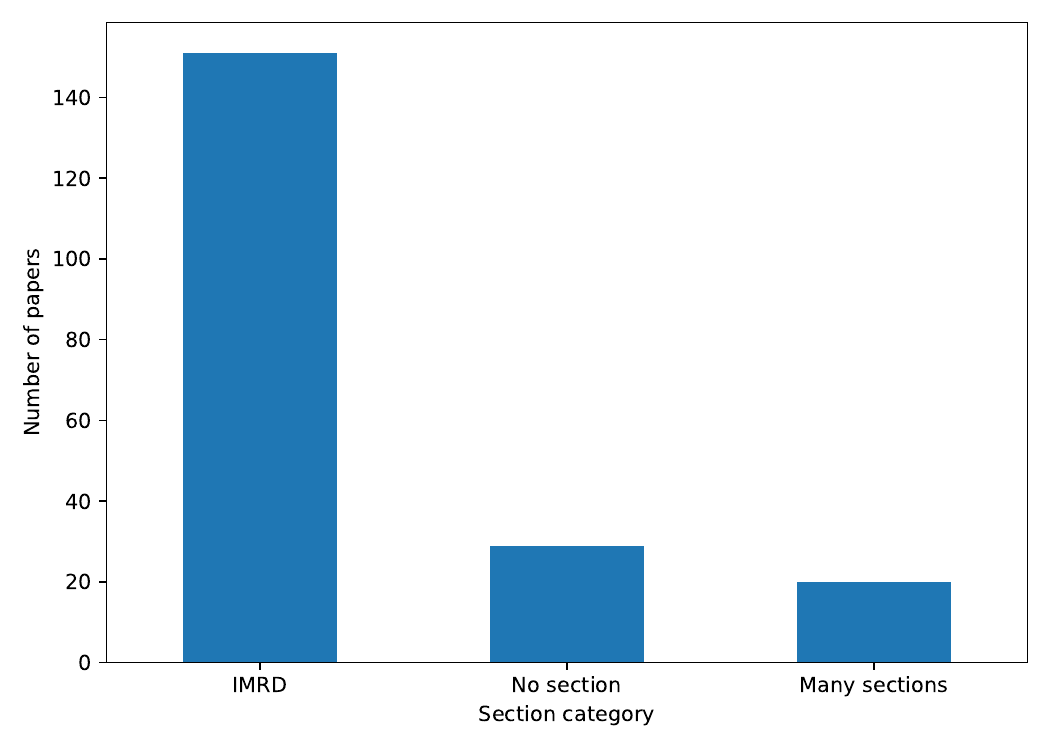}
  \caption{Distribution of section organization within the 200 papers of the dataset. IMRD is short for Introduction-Method-Results-Discussion.}
  \label{fig:section_cat}
\end{figure}

As shown in Figure~\ref{fig:section_cat}, the majority of the papers (75.5\%) follow the IMRD writing style. However, two additional writing patterns are also observed: papers that contain no explicit sectioning, such as short-format papers that may omit section headings entirely or combine Method and Result content, and papers that are divided into numerous sections whose titles do not clearly correspond to standard method or result sections. These two categories account for 14.55\% and 10\% of the analyzed papers, respectively.

In such cases, keyword-based retrieval becomes unreliable. Therefore, we apply a dense retrieval approach~\cite{karpukhin2020densepassageretrievalopendomain}, guided by concepts extracted from the abstract. This strategy is used for three of the eight papers, two of which lack section structure, while one contains number of sections.

\begin{figure}[htbp]
  \centering
  \includegraphics[width=\linewidth]{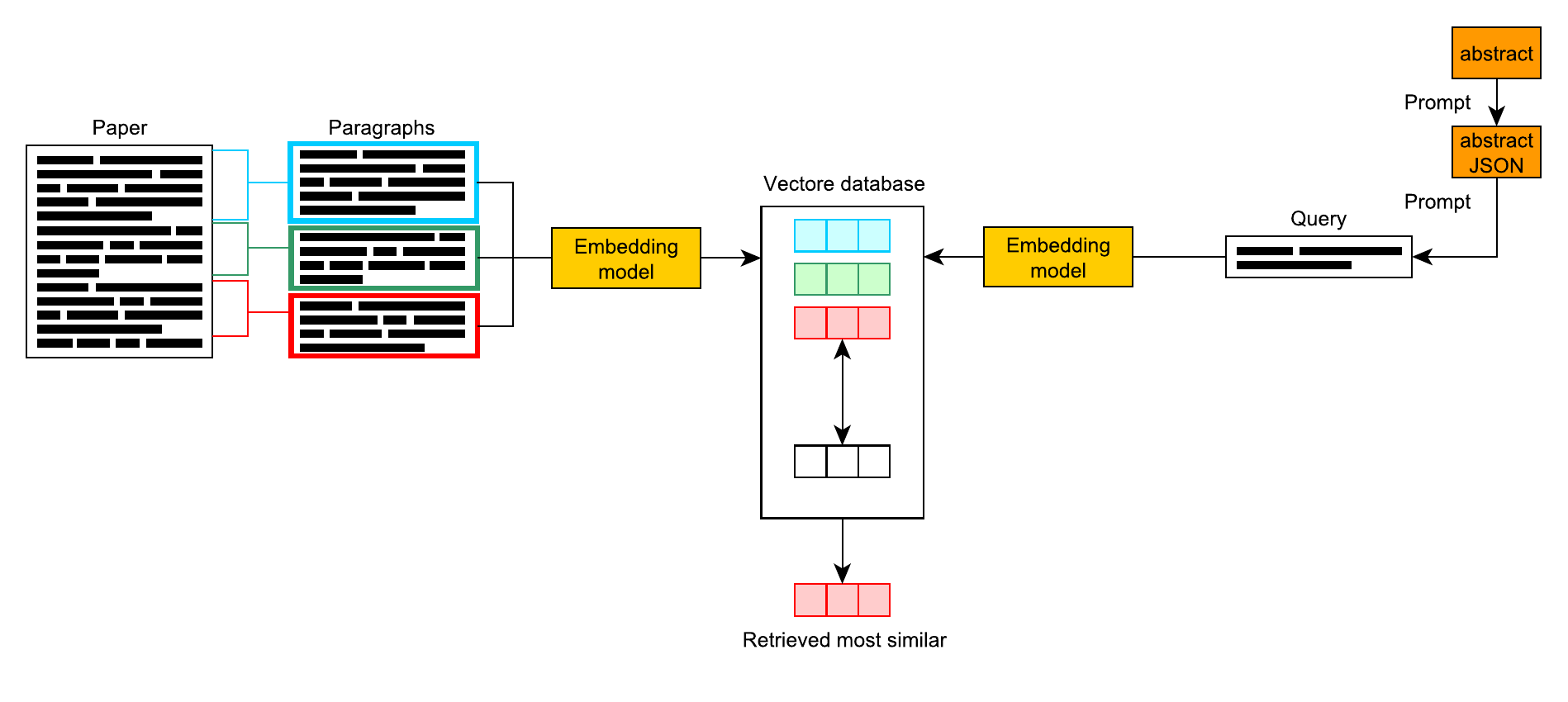}
  \caption{Schematic of LLM-based dense retrieval approach guided by concepts extracted from the abstract for identification of text passages related to the computational workflow parameters in absence of a standard IMRD structure in papers.}
  \label{fig:retrive}
\end{figure}

As shown in Figure~\ref{fig:retrive}, after converting scientific articles from PDF to Markdown using MinerU~\cite{wang2024mineruopensourcesolutionprecise}, each document is segmented into paragraph-level chunks. These chunks are then encoded into dense vector embeddings using OpenAI embedding model, text-embedding-3-large,  and stored for similarity-based retrieval.
Unlike conventional retrieval settings as mentioned in~\cite{karpukhin2020densepassageretrievalopendomain}, explicit user questions are not available. Instead, query generation is automated using LLM ($T=0$). Specifically, the abstract of each paper is provided to the LLM, which extracts key information in a structured JSON format, including \texttt{compositions}, \texttt{computational methods}, \texttt{supercell details}, and \texttt{fault details}. These keys are defined according to the workflow ontology described in Section~\ref{sec:onto_ext} and represent essential parameters required for reproducing or reusing the computational workflow.

For each extracted key-value pair, the LLM then generates one or more concise natural-language questions that serve as semantic queries. If the value of a key is empty then we ask LLM to generate the questions for the key itself.
These queries are embedded using the same embedding model as the document chunks. Dense retrieval is subsequently performed by computing vector similarity between query embeddings and paragraph embeddings, and the most relevant text chunks are retrieved for each key-value pair.
In the following, we present an example for Ref~\cite{WANG2013445} illustrating the described dense retrieval approach. First, key information is extracted from the paper abstract and normalized into a representation, where materials are mapped to the \texttt{composition} key, computational approaches are grouped under \texttt{computational\_method}, structural modeling choices are captured in \texttt{supercell\_details}, and fault-related information, including defect type, properties, and slip systems, is consolidated within \texttt{fault\_details}. Based on this structured representation, the LLM automatically generates a natural-language query for each key-value pair; in the example, a query is produced for the \texttt{computational\_method} field to retrieve methodological details relevant to defect and energy calculations. The query is then embedded and used in a dense retrieval step to identify the most semantically relevant paragraphs from the full text, which contains the detailed description of the DFT setup.

\begin{tcolorbox}[
  title={Extracted JSON representation}, 
  colback=blue!5,
  colframe=blue!60
  ]
\begin{verbatim}
{
  "composition": [
    { "material": "Mg47X1" }
  ],
  "computational_method": [
    { "method": "first-principles calculations" }
  ],
  "supercell_details": [],
  "fault_details": [
    { "type": "generalized stacking-fault energy (GSFE)" },
    { "slip_systems": "weighted consideration of each slip system" }
  ]
}

\end{verbatim}
\end{tcolorbox}

\begin{tcolorbox}[
  title={Automatically generated query}, 
  colback=blue!5,
  colframe=blue!60] 
The abstract states the use of first-principles calculations, which nearly always correspond to density functional theory (DFT). I should retrieve sections that outline the simulation setup and parameters, especially those relevant to defect and energy calculations. My goal is to extract all methodological details describing how DFT was used in the study.
\end{tcolorbox}

\begin{tcolorbox}[
  title={Retrieved paragraphs},
  colback=blue!5,
  colframe=blue!60,
  ]
\begin{Verbatim}[
  breaklines=true,
  breakanywhere=true,
  breaksymbolleft={},
  breaksymbolright={}]
"In this work, we investigate the effects of alloying elements on the mechanical properties of $\\operatorname { M g - X }$ $\\mathbf { X } =$ the doping element) alloys by calculating GSFEs in $\\{ 0 \\dot { 0 } 0 \\dot { 1 } \\} \\langle 1 1 \\bar { 2 } 0 \\rangle$ , $\\{ 1 0 \\bar { 1 } 0 \\}$ h1 1 -2 0i, $\\{ 1 0 \\bar { 1 } 1 \\}$ $\\langle 1 1 \\bar {  2 } 0 \\rangle$ and $\\{ 1 1 \\bar { 2 } 2 \\} \\quad \\langle 1 1 \\bar { 2 } 3 \\rangle$ i systems. Besides common alloying elements in $\\mathbf { M g }$ , other elements with atomic radius and/ or electron configurations different from elemental Mg were also selected, including Li, Be, Na, Al, Si, K, Ca, Ti, Cu, Zn, Ga, Ge, Sr, Zr, Ag, Cd, In, Sn, Sb, Pb and Bi (listed in atomic number order). Using the first principles data, we draw a “design map” with regard to GSFE, with the aim of providing a database for designing new $\\mathbf { M g }$ alloys. "
"The calculations here were performed in the DMOL3 code [10] based on density-functional theory, in which the Perdew and Wang [11] version of the generalized gradient approximation was employed as an exchange correlation functional. Moreover, all results were relevant for $T = 0 \\mathrm { { K } }$ . The Monkhorst–Pack mesh of $k$ -points [12] was chosen to ensure a convergence of the total energy to within $5 \\mathrm { m e V }$ atom-1 : $1 \\bar { 0 } \\times 8 \\times 1$ for the basal slip system $( \\{ 0 0 0 1 \\} \\langle 1 1 \\bar { 2 } 0 \\rangle )$ ), $8 \\times 4 \\times 1$ for the prismatic $\\dot { ( \\{ 1 0 \\bar { 1 } 0 \\} } \\quad \\langle 1 1 \\bar { 2 } 0 \\rangle )$ and pyramidal $( \\{ 1 0 \\bar { 1 } 1 \\} \\ \\dot { \\langle 1 } 1 \\bar { 2 } 0 \\rangle$ and $\\{ 1 1 \\bar { 2 } 2 \\} \\ \\langle 1 1 \\bar { 2 } 3 \\rangle )$ systems. "
\end{Verbatim}
\end{tcolorbox}

\subsection{Information extraction}
The purpose of the two previous steps is to \textit{condense} the information contained in a full article the necessary pieces. The main reason to narrow down the focus is the observation that LLMs generally perform better at extraction tasks (number of triples and fraction of correct triples) when the source text is shorter~\cite{Hsieh2024,Levy2024,Li2024a}.

\subsubsection{Structured JSON based on method section information}
Once dense chunks of methodological details have been retrieved from each paper, they are passed to Extraction Prompt~1 and processed using LLM ($T=0$), as illustrated in Figure~\ref{fig:workflow}. Papers employing the same SFE modeling strategy are subsequently grouped together, resulting in three categories: one paper using the ANNNI model, five papers using the rigid-shift approach, and two papers using the tilted-cell method. In the following, we present one representative example from the tilted-cell group. For this example, the method section is retrieved using keyword-based retrieval, and both the method-section input and the corresponding JSON output are shown.
The high-level JSON categories: \texttt{composition}, \texttt{computational\_method}, \texttt{supercell\_details}, and \texttt{fault\_details} are predefined, while the remaining keys are automatically inferred by the LLM. To facilitate comparison, key pieces of information in both the input text and the generated JSON output are highlighted using consistent colors.
Overall, the LLM accurately captures the core components of the calculation workflow. Information highlighted in blue in the input, corresponding to the exchange-correlation functional (GGA-PW91), is correctly identified and mapped to the \texttt{exchange\_correlation\_functional} field in the output. Similarly, the green-highlighted supercell information, including the basal slip plane \(\{0001\}\), supercell dimensions, and number of atoms, is correctly extracted and represented under \texttt{supercell\_details}. The red-highlighted $k$-point mesh for the basal slip plane is also accurately transferred to the output, demonstrating reliable extraction of numerical simulation parameters. In addition, the orange-highlighted supercell tiling technique is correctly recognized as the method used to compute the generalized stacking fault (GSF) energy surface and is appropriately placed under \texttt{fault\_details}.
This indicates that the LLM performs well in extracting individual parameters and in capturing the relational dependencies between different parameters, such as the association between slip planes and their corresponding $k$-point meshes.
Furthermore, the output includes some information that is not strictly necessary for defining the computational workflow. For example, explanatory statements regarding the physical significance of the $\gamma$-surface and its relevance to dislocation core analysis, while scientifically meaningful, are not essential for reproducing the computational workflow itself. Conversely, certain workflow-relevant details, such as the explicit construction procedure of the tilted cell or constraints applied during relaxation, are not present in the input text and therefore cannot be recovered.
\vspace{1em}

\noindent
\begin{minipage}[t]{0.48\textwidth}
\textbf{Input:}
{\renewcommand{\baselinestretch}{0.90}
\begin{lstlisting}[style=method]
2 Computational details

In this work, all calculations were implemented in the Vienna Ab initio Simulation Package (VASP) [19]. The (*@\textcolor{blue}{Perdew Wang (PW91) version of the generalized gradient approximation (GGA)}@*)[20] is used to treat exchangecorrelation functional and the projector augmented wave (PAW) method [21] has been used in the present work. The cutoff energy of plane wave was chosen at 350 eV. The total-energy calculation was performed until the total energy changes within $1 0 ^ { - 6 }$ eV/atom and the Hellmann-Feynman force on all atomic sites was less than $1 0 ^ { - 2 } \ \mathrm { e V / \AA }$ . 

According to the different slip planes, the shapes and sizes of the supercells used in the present study are not the same. For the (*@\textcolor{green}{basal slip plane $\{ 0 0 0 1 \}$}@*) and pyramidal slip plane $\{ 1 0 \overline { { 1 } } 1 \}$ , (*@\textcolor{green}{a $1 \times 1 \times 6$ periodic supercell with 12 atoms}@*) was used (except 13 atoms for $E$ type stacking fault ), whereas a $1 \times 1 \times 4$ supercell with 16 atoms (a unit-cell contains 4 atoms) was used for the prismatic slip plane $\{ 1 0 \overline { { 1 } } 0 \} $ , $\{ 1 1 \overline { { 2 } } 0 \}$ and pyramidal slip plane $\{ 1 1 \overline { { 2 2 } } \}$ . The Brillouin zone integrations were performed with Gamma centered Monkhorst-Pack grids [22]. Based on the different shapes and sizes of supercells, different $k$ -points were used to calculate the total energy for different slip planes. After enough tests, (*@\textcolor{red}{$1 5 \times 1 5 \times 3$ for basal slip plane supercell}@*), $1 5 \times 1 1 \times 5$ for prismatic slip plane $\{ 1 0 \overline { { 1 0 } } \}$ supercell, $1 3 \times 1 1 \times 5$ for prismatic slip plane $\{ 1 1 \overline { { 2 } } 0 \}$ supercell, $1 3 \times 9 \times 7$ for pyramidal slip plane $\lbrace 1 0 \rbrace 1 \rbrace$ supercell and $1 3 \times 1 1 \times 5$ for pyramidal slip plane $\{ 1 1 2 2 \}$ supercell were used. The convergence tests with respect to these parameters show that the error bar for the total energy is less than 0.1 meV/atom. 

We used a (*@\textcolor{orange}{supercell tiling technique}@*) to calculate the generalized stacking fault (GSF) energy surface ($\gamma$-surface), which has been described by Yan and his coworkers~[4]. The GSF energy surface provides a comprehensive description of stacking faults~[23]. In particular, the intrinsic stacking fault width in FCC alloys has been shown to depend quantitatively on the generalized stacking fault energy curve~[24]. According to Ref.~[25], a ``proper'' $\gamma$-surface is essential for analyzing dislocation cores and identifying key deformation properties within the scope of the P-N model, thereby improving the understanding of mechanical properties. Consequently, the GSF energy surface is a central focus of the present investigation.


3 Results and discussion 
\end{lstlisting}}
\end{minipage}
\hfill
\begin{minipage}[t]{0.48\textwidth}
\textbf{Output (JSON):}

\begin{lstlisting}[style=jsonbox]
{
  "compositions": [
    {
      "composition": "Not specified",
      "computational_method": {
        "software": "VASP",
        "exchange_correlation_functional": "<blue>GGA (PW91)</blue>",
        "pseudopotential": "PAW",
        "plane_wave_cutoff_energy": {
          "value": 350,
          "unit": "eV"
        },
        "energy_convergence_criterion": {
          "value": 1e-06,
          "unit": "eV/atom"
        },
        "force_convergence_criterion": {
          "value": 0.01,
          "unit": "eV/A"
        },
        "brillouin_zone_integration": "Gamma centered Monkhorst-Pack grids",
        "total_energy_error_bar": {
          "value": 0.1,
          "unit": "meV/atom"
        }
      },
      "supercell_details": [
        {
          "slip_plane": "<green>{0001} (basal)</green>",
          "supercell_size": "<green>1 x 1 x 6</green>",
          "number_of_atoms": <green>12</green>,
          "k_points": "<red>15 x 15 x 3</red>"
        },
        {
          "slip_plane": "{10(-1)1} (pyramidal)",
          "supercell_size": "1 x 1 x 6",
          "number_of_atoms": 12,
          "k_points": "13 x 9 x 7"
        },
        {
          "slip_plane": "{10(-1)1} (pyramidal, E-type stacking fault)",
          "supercell_size": "1 x 1 x 6",
          "number_of_atoms": 13,
          "k_points": "13 x 9 x 7"
        },
        {
          "slip_plane": "{10(-1)0} (prismatic)",
          "supercell_size": "1 x 1 x 4",
          "number_of_atoms": 16,
          "k_points": "15 x 11 x 5"
        },
        {
          "slip_plane": "{11(-2)0} (prismatic)",
          "supercell_size": "1 x 1 x 4",
          "number_of_atoms": 16,
          "k_points": "13 x 11 x 5"
        },
        {
          "slip_plane": "{11(-2)2} (pyramidal)",
          "supercell_size": "1 x 1 x 4",
          "number_of_atoms": 16,
          "k_points": "13 x 11 x 5"
        }
      ],
      "fault_details": {
        "fault_type": "Generalized stacking fault (GSF) energy surface",
        "calculation_method": "<orange>Supercell tilting technique (Yan et al.)</orange>",
        "purpose": [
          "Comprehensive description of stacking faults",
          "Analysis of dislocation core",
          "Detection of deformation properties in the P-N model",
          "Understanding mechanical properties"
        ]
      }
    }
  ]
}
\end{lstlisting}
\end{minipage}

\subsubsection{JSON creation of the SFE table}
For each SFE-relevant table, the extracted table content is combined with the corresponding method-section JSON keys and processed using Extraction Prompt~2 with an LLM ($T=0$), yielding an enriched JSON representation of the tabular data as shown in Figure~\ref{fig:workflow}. In the following, we present one representative example from the tilted-cell group, showing both the table input and the generated JSON output.
The table caption and footnote play a critical role in providing contextual information that is not contained within the numerical entries alone. In this example, the caption specifies that the reported values correspond to stacking fault energies of basal plane stacking faults and provides the physical unit (mJ/m$^{2}$), while the footnotes identify references associated with different computational and experimental sources. This information is successfully incorporated into the JSON output, where the unit is explicitly assigned to each extracted stacking fault energy value, and the general fault context is summarized under the \texttt{fault\_details} category.
The table structure obtained from the MinerU output enables the LLM to correctly interpret the column headers as stacking fault labels (I1, I2, E, and T2) and the rows as different data sources. Guided by the prompt design, which restricts extraction to values corresponding to the \emph{present work}, only the numerical entries associated with the row labeled ``The present work GGA-PAW'' are included in the structured output. As a result, values from other computational approaches (GGA-PBE, LDA) and experimental references are correctly ignored, demonstrating that the prompt effectively filters table content based on relevance to the target calculation workflow.
Keys defined in the method-section JSON, such as \texttt{exchange\_correlation\_functional}, are reused during table extraction to ensure consistency between methodological metadata and tabulated results. In this example, the exchange-correlation functional GGA-PAW, identified earlier in the method section, is correctly propagated into the table-level JSON. Additionally, keys such as \texttt{fault\_type} are populated based on information inferred from the table caption. While the prompt allows the LLM to introduce new keys when relevant information is not present in the method-section JSON, the overall structure still adheres to the four predefined high-level categories: \texttt{composition}, \texttt{computational\_method}, \texttt{supercell\_details}, and \texttt{fault\_details}.
Despite the overall correctness of the extraction, certain semantic ambiguities are observed. For instance, the value ``basal plane stacking faults'' is assigned to the \texttt{fault\_type} field, whereas from a crystallographic perspective, ``basal'' refers to the slip plane rather than the fault type itself. Similarly, labels such as I1, I2, E, and T2, which are encoded as \texttt{fault\_label} in the output, represent distinct stacking fault types. This indicates that while the LLM accurately captures and structures the numerical data and their associations, the distinction between slip plane descriptors and fault-type nomenclature is not always resolved correctly without additional domain-specific constraints.

\vspace{1em}

\noindent
\begin{minipage}[t]{0.48\textwidth}
\textbf{Input:}
\begin{lstlisting}[style=inputbox]
<table>
  <tr>
    <td><green>I1</green></td>
    <td>I2</td>
    <td>E</td>
    <td>T2</td>
  </tr>
  <tr>
    <td><blue>The present work GGA-PAW</blue></td>
    <td><red>16</red></td>
    <td>34</td>
    <td>59</td>
    <td>38</td>
  </tr>
  <tr>
    <td>Calculation GGA-PBEa</td>
    <td>18</td>
    <td>36</td>
    <td>58</td>
    <td>40</td>
  </tr>
  <tr>
    <td>Calculation LDAb</td>
    <td>21</td>
    <td>44</td>
    <td>69</td>
    <td>51</td>
  </tr>
  <tr>
    <td>Experimentc</td>
    <td></td>
    <td>&lt;50; &gt;100</td>
    <td></td>
    <td></td>
  </tr>
</table>

Table 2: The stacking fault energies of the four basal plane stacking faults <orange>(mJ/m^2)</orange>.
Fotenote: $^ { a }$ Reference [2]. $^ { b }$ Reference [3]. $_ c$ References [6,30,36].

\end{lstlisting}
\end{minipage}
\hfill
\begin{minipage}[t]{0.48\textwidth}
\textbf{Output (JSON):}

\begin{lstlisting}[style=jsonbox]
{
  "compositions": [
    {
      "composition": "Not specified",
      "computational_method": {
        "exchange_correlation_functional": "<blue>GGA-PAW</blue>"
      },
      "fault_details": {
        "fault_type": "basal plane stacking faults",
        "purpose": [
          "stacking fault energy calculation"
        ],
        "stacking_fault_energies": [
          {
            "fault_label": "<green>I1</green>",
            "value": <red>16</red>,
            "unit": "<orange>mJ/m^2</orange>"
          },
          {
            "fault_label": "I2",
            "value": 34,
            "unit": "mJ/m^2"
          },
          {
            "fault_label": "E",
            "value": 59,
            "unit": "mJ/m^2"
          },
          {
            "fault_label": "T2",
            "value": 38,
            "unit": "mJ/m^2"
          }
        ]
      }
    }
  ]
}
\end{lstlisting}
\end{minipage}

\subsubsection{JSON unification}
\begin{figure}[htbp]
  \centering
  \includegraphics[width=.5\linewidth]{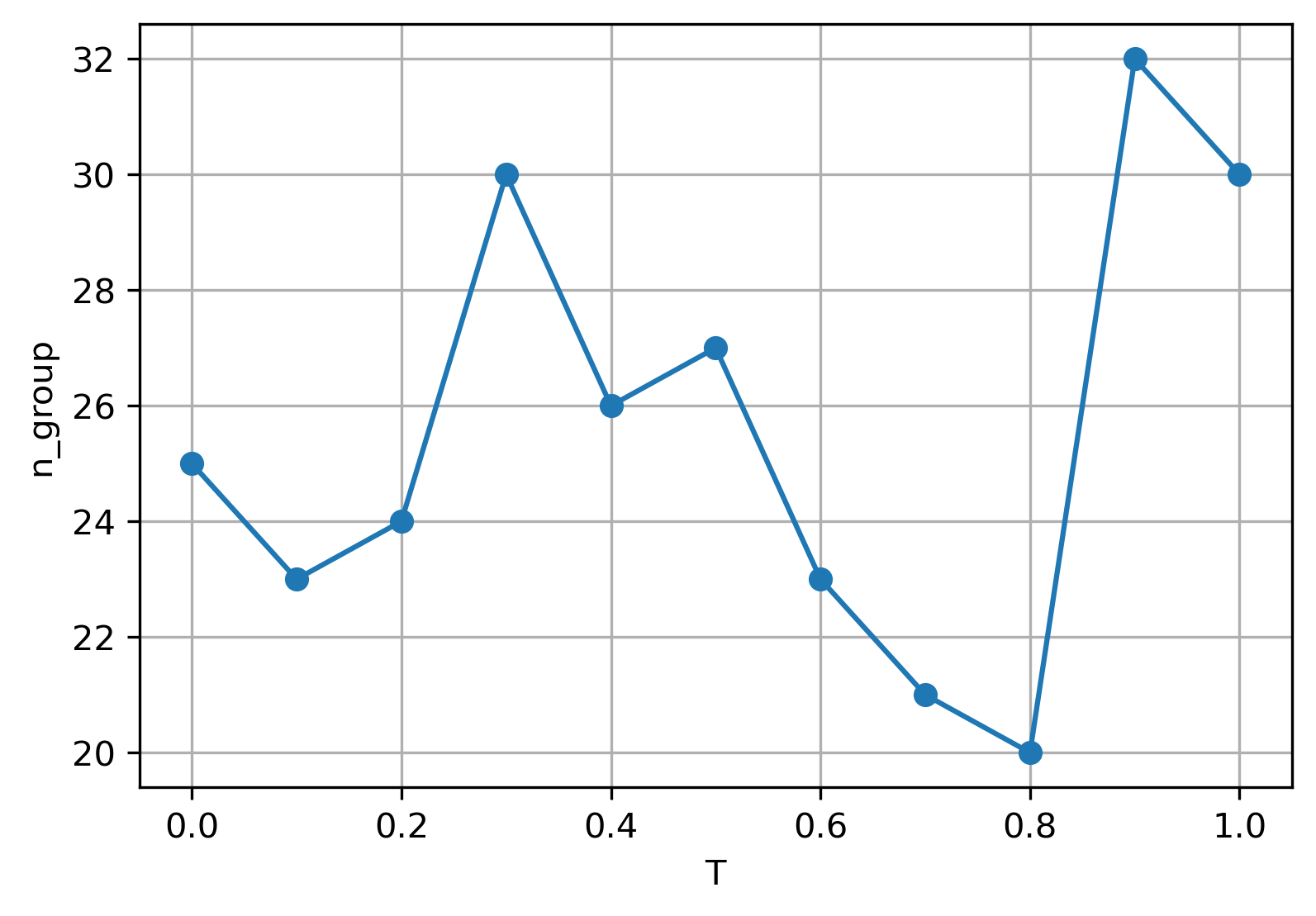}
  \caption{Number of groups as a function of different LLM temperatures $T$.}
  \label{fig:temperature_grouping}
\end{figure}

Because the method sections and tables of each paper are extracted independently, the resulting JSON representations do not follow a consistent schema. Although keys from the method-section JSON are reused during table-level JSON creation, the LLM is explicitly allowed to introduce new keys when relevant information is not present in the method section. Consequently, the same underlying concept may appear under different keys across papers, simply because it is described differently in the original text. This variability necessitates the construction of a unified, canonical schema.
To create such a unified schema, we first extract all unique key-value pairs from each JSON pair generated from the method sections and the corresponding tables. This step removes redundant keys and provides a comprehensive set of candidate concepts.
Next, we identify semantic similarities among these key-value pairs, as illustrated in Listing~\ref{lst:group_keys}, where conceptually equivalent terms such as ``slip plane'' and ``fault plane'' must be grouped together despite their different representation.
Semantic grouping is performed using an LLM with varying temperature values between 0 and 1.
We also tested word embeddings for this task but they led to ambiguous classification.
For our dataset we observed consistent, good results using an LLM.
As shown in Figure~\ref{fig:temperature_grouping}, the number of resulting semantic groups exhibits only moderate variation across this temperature range, indicating that the grouping process is relatively robust w.r.t. different temperatures $T$. Based on this analysis, LLM with a temperature of 0.3 is selected for semantic grouping, as it yields a balanced number of groups while avoiding both over-fragmentation and excessive merging of distinct concepts.

After semantic grouping, we use the LLM ($T=0.5$) to generate a canonical schema for each group. The purpose of this step is to derive a minimal, yet comprehensive representation that captures the essential semantic content shared across heterogeneous key-value pairs within a group.
As illustrated in Listing~\ref{lst:group_keys}, a single semantic group may contain a wide variety of surface forms describing closely related crystallographic concepts, including slip planes, fault planes, slip directions, and combined slip-system notations. These concepts are expressed using different terminologies, levels of specificity, and notational conventions across papers (\texttt{slip\_plane}, \texttt{fault\_plane}, \texttt{plane}, or combined plane-direction expressions). Generating a canonical schema therefore requires abstraction beyond direct key matching, while still preserving domain-specific distinctions.
A moderate temperature ($T=0.5$) is used for the schema-generation step. As demonstrated in the Appendix, the number of canonical keys remains largely invariant with respect to $T$. Consequently, the $T$ is chosen based on the semantic clarity and interpretability of the resulting canonical schema rather than on quantitative differences in schema structure.
The resulting canonical schema, shown in Listing~\ref{lst:cano_keys}, unifies the diverse representations observed in Listing~1 into a small set of well-defined keys, such as \texttt{plane\_family}, \texttt{plane\_miller}, \texttt{direction}, and \texttt{slip\_system}. This schema preserves the essential crystallographic information needed for workflow interpretation while eliminating redundancy and inconsistency across the original JSON representations.

\vspace{1em}
\noindent
\begin{minipage}[t]{0.48\textwidth}
{\renewcommand{\baselinestretch}{0.77}
\begin{lstlisting}[style=jsonbox, caption={Grouping keys.}, label={lst:group_keys}]
{
  "group_name": "Slip Planes, Slip Systems, and Directions",
  "key_value_pairs": {
    "slip_plane": [
      "basal {0001}",
      "pyramidal {10-11}",
      "pyramidal {10-11} (E type stacking fault)",
      "prismatic {10-10}",
      "prismatic {11-20}",
      "pyramidal {11-22}"
    ],
    "plane": [
      "Basal",
      "Prism I",
      "Prism II",
      "Pyramidal I",
      "Pyramidal II",
      "{0001}",
      "{10-10}",
      "{10-11}",
      "{11-22}"
    ],
    "slip_systems.slip_plane": [
      "{0001}",
      "{1-100}",
      "{11-22}"
    ],
    "slip_systems.slip_direction": [
      "[1-100]",
      "[11-20]",
      "[11-23]"
    ],
    "slip_systems.": [
      "{0001}<11-20>",
      "{10-10}<11-20>",
      "{10-11}<11-20>",
      "{11-22}<11-23>"
    ],
    "slip_system": [
      "{10-11} <11-23>",
      "(0 0 0 1)[1 1 2 0] (perfect dislocation, PD)",
      "{11 00}[1 2 0] (perfect dislocation, PD)",
      "(0 0 0 1)[1 1 0 0] (partial or Shockley dislocation)",
      "{11-22}<11-23>",
      "{10-11}<11-20>",
      "{11-20}<10-11>",
      "{10-10}<11-20>",
      "{11-20}<11-20>",
      "{11-23}<11-23>",
      "{0001}<11-20>",
      "(0001)[11-20]",
      "(0001)[1-120]",
      "{11-100}[1-120]",
      "(0001)[1-100]"
    ],
    "slip_systems.plane": [
      "{0001}",
      "{10-10}",
      "{10-11}",
      "{11-22}"
    ],
    "slip_systems.direction": [
      "<11-20>",
      "<11-23>"
    ],
    "direction": [
      "<11-20>",
      "<11-23>"
    ],
    "directions.": [
      "<00 01>",
      "<1 00>"
    ],
    "fault_plane": [
      "{11-22}(11-23)",
      "{0001}(1-100)",
      "{1-100}(11-20)",
      "Basal",
      "Pyramidal I-N",
      "Prism II",
      "Pyramidal I-W",
      "Prism I-W"
    ],
    "fault_plane_direction": [
      "{10-11}<11-20>",
      "{10-10}<11-20>",
      "{10-11}<11-23>",
      "{0001}<11-20>",
      "{11-22}<11-23>",
      "l2"
    ]
  }
}
\end{lstlisting}}
\end{minipage}
\hfill
\vspace{1cm}
\begin{minipage}[t]{0.48\textwidth}
\begin{lstlisting}[style=jsonbox,caption={Canonical keys.}, label={lst:cano_keys}]
"canonical_keys": {
  "plane_family":
    "Crystallographic family of the slip or fault plane (e.g., basal, prismatic, pyramidal)",
  "plane_miller":
    "Miller or Miller-Bravais indices of the plane (e.g., {0001}, {10-10})",
  "direction":
    "Crystallographic direction or direction family (e.g., <11-20>, [1-100])",
  "slip_system":
    "Combined slip system notation (plane + direction, e.g., {0001}<11-20>)",
}
\end{lstlisting}
\end{minipage}

\subsubsection{Canonical JSON generation}
Once a canonical schema is established, it is used to homogenize the previously heterogeneous JSON representations (keys) generated independently from method sections and tables. This step is necessary because, prior to canonicalization, semantically equivalent information may appear under different keys or structures across papers, preventing direct comparison, aggregation, or reuse.
To achieve this homogenization, the canonical schema is applied to the original Table and the corresponding method section of each paper. Using Extraction Prompt~3 with the LLM ($T=0.5$), we map and extract relevant information from both sources into the predefined canonical fields. This process explicitly links paper-specific representations back to a unified schema, thereby resolving inconsistencies introduced during independent extraction.

The outcome of this process is a homogeneous representation in which each numerical table entry is paired with its complete methodological context. From a workflow perspective, the resulting records can be interpreted as structured input-output pairs, analogous to those produced by a computational workflow: the \textit{method} block encodes the computational inputs and settings, while the \textit{data\_point} block represents the corresponding calculated result.

An example of such a homogenized representation is shown below for Ref.~\cite{wen2009systematic}. Here, methodological details extracted from the method section such as DFT software, exchange-correlation functional, supercell size, and $k$-point sampling are combined with a single table entry describing a basal SFE. This unified JSON representation explicitly links the SFE value to its full computational provenance through the canonical schema, enabling comparison and reuse across different papers and workflows.

\vspace{1em}

\noindent
\begin{minipage}[t]{0.40\textwidth}
\textbf{Output (JSON):}
\begin{lstlisting}[style=jsonbox]
{
  "method": {
    "dft_software": "VASP",
    "xc_functional": "GGA-PW91",
    "pseudopotential": "PAW",
    "plane_wave_cutoff_energy": "350 eV",
    "energy_convergence_criterion": "1e-6 eV/atom",
    "force_convergence_criterion": "1e-2 eV/A",
    "k_point_grid": "15x15x3",
    "supercell_size": {
      "slip_plane": "{0001}",
      "supercell_size": "1x1x6",
      "number_of_atoms": "12"
    },
    "k_point_grid_type": "Monkhorst-Pack",
    "k_point_grid_centering": "Gamma",
    "total_energy_error_bar": "0.1 meV/atom",
    "calculation_method":
      "supercell tilting technique to calculate the generalized stacking fault (GSF) energy surface",
    "note":
      "GSF energy surface calculation method described by Yan et al. [4]; GSF energy surface is the focus of the present investigation."
  },

  "data_point": {
    "composition": "Mg",
    "fault_type": "basal stacking fault",
    "fault_label": "I1",
    "fault_name": "I1 stacking fault",
    "stacking_fault_energy": 16,
    "stacking_fault_energy_unit": "mJ/m^2"
  }
}

\end{lstlisting}
\end{minipage}

\subsection{Ontology alignment}
So far, we have relied exclusively on information extracted from method sections and tables using a prompting strategy applied across the corpus. While this approach enables the construction of a literature-derived canonical schema, it does not yet provide a formally grounded semantic representation that allows integration, querying, or interoperability beyond the scope of this study. To address this limitation, we map the literature-based canonical schema to established ontological concepts as introduced in Section~\ref{ssec:onto_sfe}.
Specifically, each canonical key-value pair is aligned with corresponding concepts defined in three domain ontologies: CMSO~\cite{CMSO}, ASMO~\cite{ASMO}, and PLDO~\cite{PLDO} as it has been shown in Figure~\ref{fig:SFE_workflow_onto}.
In the following, we present one representative example illustrating the ontology-mapping process for a single data point extracted from Ref~\cite{wen2009systematic}. The input (Listing~\ref{lst:canonical_input_example}) corresponds to a homogenized, literature-derived JSON representation that combines methodological information with a single table-derived stacking fault energy value. The output (Listing~\ref{lst:ontology_output_example}) shows the result of mapping this canonical representation to ontology-aligned entities and relations.
In this example, canonical fields such as \texttt{composition}, \texttt{supercell\_size}, and \texttt{stacking\_fault\_energy} are translated into ontological instances. CMSO~\cite{CMSO} concepts are used to represent the computational samples, including chemical species, crystal structure, and simulation cell repetitions as colored in blue. PLDO~\cite{PLDO} concepts are employed to describe the stacking fault, including its type (I1), fault plane, and crystallographic label as colored in green. ASMO~\cite{ASMO} concepts capture the computational method, including the use of density functional theory, the VASP software package, the exchange-correlation functional, plane-wave cutoff energy, and $k$-point sampling as colored in orange.
\vspace{1em}

\noindent
\begin{minipage}[t]{0.48\textwidth}
\textbf{Input (canonical JSON):}
{\renewcommand{\baselinestretch}{0.80}
\begin{lstlisting}[style=inputbox, label={lst:canonical_input_example}
]]
{
  "method": {
    "dft_software": "<orange>VASP</orange>",
    "xc_functional": "<orange>GGA-PW91</orange>",
    "pseudopotential": "<orange>PAW</orange>",
    "plane_wave_cutoff_energy": "<orange>350 eV</orange>",
    "energy_convergence_criterion": "<orange>1e-6 eV/atom</orange>",
    "force_convergence_criterion": "<orange>1e-2 eV/A</orange>",
    "k_point_grid": "<orange>15x15x3</orange>",
    "supercell_size": {
      "slip_plane": "<green>{0001}</green>",
      "supercell_size": "<blue>1x1x6</blue>",
      "number_of_atoms": "<blue>12</blue>"
    },
    "k_point_grid_type": "<orange>Monkhorst-Pack</orange>",
    "k_point_grid_centering": "<orange>Gamma</orange>",
    "total_energy_error_bar": "<orange>0.1 meV/atom</orange>",
    "calculation_method":
      "<orange>supercell tilting technique to calculate the generalized stacking fault (GSF) energy surface</orange>",
    "note":
      "<orange>GSF energy surface calculation method described by Yan et al. [4]</orange>"
  },

  "data_point": {
    "composition": "<blue>Mg</blue>",
    "fault_type": "<green>basal stacking fault</green>",
    "fault_label": "<green>I1</green>",
    "fault_name": "<green>I1 stacking fault</green>",
    "stacking_fault_energy": "<green>16</green>",
    "stacking_fault_energy_unit": "<green>mJ/m^2</green>"
  }
}

\end{lstlisting}}
\end{minipage}
\hfill
\begin{minipage}[t]{0.48\textwidth}
\textbf{Output (ontology-aligned JSON):}
{\renewcommand{\baselinestretch}{0.60}
\begin{lstlisting}[style=jsonbox, label={lst:ontology_output_example}]
{
 "ComputationalSample_1": {
    "ChemicalSpecies": {
      "ChemicalElement": "<blue>Mg</blue>",
      "SubstitutionalSolute": null,
      "hasSoluteConcentration": 0.0
    },
    "ChemicalComposition": "<blue>Mg12</blue>",
    "Material": "<blue>metal</blue>",
    "CrystalStructure": "<blue>hcp</blue>",
    "SimulationCell": {
      "SimulationCellVector": {
        "hasComponent_x": null,
        "hasComponent_y": null,
        "hasComponent_z": null
      },
      "SimulationCellLength": {
        "hasLength_x": null,
        "hasLength_y": null,
        "hasLength_z": null
      },
      "SimulationCellAngle": {
        "hasAngle_alpha": null,
        "hasAngle_beta": null,
        "hasAngle_gamma": null
      },
      "hasRepetition_x": <blue>1</blue>,
      "hasRepetition_y": <blue>1</blue>,
      "hasRepetition_z": <blue>6</blue>,
      "hasUnitCellType": null,
      "hasVacuumLength": null,
      "hasVacuumDirection": null
    }},
  "ComputationalSample_2": {
    "ChemicalSpecies": {
      "ChemicalElement": "<blue>Mg</blue>",
      "SubstitutionalSolute": null,
      "hasSoluteConcentration": 0.0
    },
    "ChemicalComposition": "<blue>Mg12</blue>",
    "Material": "<blue>metal</blue>",
    "CrystalStructure": "<blue>hcp</blue>",
    "SimulationCell": {
      "SimulationCellVector": {
        "hasComponent_x": null,
        "hasComponent_y": null,
        "hasComponent_z": null
      },
      "SimulationCellLength": {
        "hasLength_x": null,
        "hasLength_y": null,
        "hasLength_z": null
      },
      "SimulationCellAngle": {
        "hasAngle_alpha": null,
        "hasAngle_beta": null,
        "hasAngle_gamma": null
      },
      "hasRepetition_x": <blue>1</blue>,
      "hasRepetition_y": <blue>1</blue>,
      "hasRepetition_z": <blue>6</blue>,
      "hasUnitCellType": null,
      "hasVacuumLength": null,
      "hasVacuumDirection": null
    },
    "StackingFault": {
      "Type": "<green>I1StackingFault</green>",
      "hasDisplacementVector": null,
      "hasSFplane": "<green>(0001)</green>",
      "hasStackingSequence": null,
      "SF_hcp_label": "<green>basal</green>"
    }},
  "<green>I1StackingFaultEnergy</green>": {
    "value": <green>16</green>,
    "unit": "<green>http://qudt.org/vocab/unit/MilliJ-PER-M2</green>",
    "label": "<green>stable</green>"
  },
  "<orange>ComputationalMethod</orange>": {
    "Simulation": "<orange>Tilted Cell</orange>",
    "Type": "<orange>DensityFunctionalTheory</orange>",
    "SoftwareAgent": "<orange>VASP</orange>",
    "XC_EnergyFunctional": "<orange>GGA-PW91</orange>",
    "EnergyCutoff": {
      "value": <orange>350.0</orange>,
      "unit": "<orange>http://qudt.org/vocab/unit/EV</orange>"
    },
    "KpointMesh": {
      "Type": "<orange>MonkhorstPackKPointMesh</orange>",
      "value": <orange>[15,15,3]</orange>
    },
    "InputParameter": {
      "Pseudopotential": {
        "label": "<orange>PAW</orange>"
      },
      "EnergyConvergence": {
        "value": <orange>0.000001</orange>,
        "unit": "<orange>http://qudt.org/vocab/unit/EV</orange>",
        "label": "<orange>energy_convergence_for_relaxation_per_atom</orange>"
      },
      "ForceConvergence": {
        "value": <orange>0.01</orange>,
        "unit": "<orange>http://qudt.org/vocab/unit/EV-PER-ANGSTROM</orange>",
        "label": "<orange>force_convergence_for_relaxation</orange>"
      }
    }
  }
}

\end{lstlisting}}
\end{minipage}

The canonical JSON object represents the raw extracted information from scientific papers.
It contains two main components: a method block, which describes the computational workflow used to calculate SFE, and a data point block, which contains individual SFE values together with metadata such as the fault type and material system. 
The ontology-aligned JSON object constitutes a reorganization of same information into a knowledge graph-compatible structure that can be translated to triples.
This separates concepts into semantically-defined entities with all implied/derived relationships such as units for ComputationalSample, ComputationalMethod, StackingFault, and StackingFaultEnergy, etc.

However, this canonical JSON representation has the minimum ontology classes, and to truly make it ready for knowledge graph creation, it needs to be modified. These changes are primarily concerned with making relationships explicit, separating concerns between samples, workflows, and operations, and ensuring that all computed properties and transformations are formally linked rather than implied.

In its current form, the canonical JSON describes the correct scientific content: HCP Mg, an I1 basal stacking fault, and a DFT tilted-cell calculation, but it does so in an object-centric, ontology-flavored way where key entities exist as isolated blocks. Computational samples, stacking fault energies, and computational methods are all present, yet the semantic relationships between them are implicit. For example, the stacking fault energy appears as a standalone object rather than being explicitly associated with the defected sample, and the computational method is not embedded within a workflow that consumes inputs and produces outputs. This limits the ability to automatically construct a knowledge graph without additional interpretation logic.

The revised representation restructures the same information into a schema that is explicitly relational and workflow-aware. Computational samples are defined as identifiable entities with stable IDs, allowing both pristine and defected structures to coexist and be referenced unambiguously. Defects such as stacking faults are modeled as structured components of a sample, and defect complexes are explicitly declared, enabling extension to multi-defect systems. This makes each sample a well-defined node in a future knowledge graph.

A major change is the introduction of an explicit workflow layer. Instead of a standalone computational method, the calculation is represented as a workflow that specifies its inputs, computational parameters, software, and outputs. Calculated properties, such as the I1 stacking fault energy, are produced by the workflow and explicitly associated with the appropriate sample. This creates clear provenance edges in the graph: which workflow produced which property, using which inputs.

Another important change is the explicit modeling of structural transformations. The generation of the stacking fault is no longer implicit; it is captured as a shear operation that transforms the bulk Mg sample into the stacking-faulted sample. This makes the causal relationship between structures machine-readable and supports graph queries such as defect derivation paths or transformation histories.

After creating the ontology-aligned JSON, we used its JSON keys as the reference schema for our reference dataset. When comparing the methods sections across all papers, we observed that the extracted methodological parameters were consistent and correctly identified for all studies. As a result, the evaluation focuses primarily on the extraction of numerical data from tables rather than on methodological descriptions.

For evaluation, a data point is defined as a single SFE value together with all its associated dependencies as reported in the table, such as composition, fault type, slip system, or calculation mode. Based on this definition, we assessed the extraction performance on a paper-by-paper basis.

For Ref.~\cite{DOU2015405}, the table contains 30 data points, of which 25 were extracted correctly. The five incorrect entries correspond to pyramidal stacking fault configurations, where the SFE type was incorrectly classified as unstable rather than stable.

Ref.~\cite{dou2019generalized} contains two tables. The first table includes 75 data points; the model extracted 72 of these, with 66 correct entries. The primary source of error was incorrect composition parsing, particularly cases where compositions reported as Mg$_{56}$ were extracted simply as Mg. The second table in this paper contains 51 data points, all of which were extracted correctly.

In Ref.~\cite{jiang2025first}, two tables were evaluated. The first table contains 115 data points, of which 37 were extracted incorrectly. The dominant error in this table was a crystallographic misinterpretation, where the slip direction $\langle 11\bar{2}3 \rangle$ was incorrectly identified as the slip plane. The second table in this paper contains 65 data points, all of which were extracted correctly.

Ref.~\cite{ctx5177087330006471} contains a single table with four data points, all of which were extracted correctly.

Ref.~\cite{wen2009systematic} includes two tables, one containing four data points and the other containing six data points. All entries in both tables were extracted correctly.

For Ref.~\cite{yin2017comprehensive}, the table contains 45 data points, all of which were extracted correctly. However, five additional data points described only in the table caption, rather than in the tabulated body, were not extracted and therefore count as missed entries.

Finally, Ref.~\cite{HU20131136} contains two tables. The first table includes 112 data points, all of which were extracted correctly. The second table contains 24 data points, and all were also extracted correctly.

\subsection{Knowledge graph query}
Table~\ref{tab:sfe_mg} summarizes reported values of the intrinsic I2 SFE for basal $\{0001\}$ planes in pure Mg obtained from different theoretical studies. The query of interest is specifically the I2 stacking fault on the basal plane of Mg, which is relevant for basal slip processes.

Multiple entries from Ref.~\cite{DOU2015405} are listed because the authors employed different structural relaxation schemes while keeping the same $k$-point mesh and plane-wave cutoff. Despite these methodological variations, the calculated SFE values are highly consistent, lying within a narrow range of 35--36~mJ~m$^{-2}$. This indicates that the basal I2 SFE in Mg is relatively insensitive to the details of the relaxation procedure within the rigid-shift approach.

Across different computational methods, the reported SFE values are broadly comparable once unit conversions are taken into account. For instance, Ref.~\cite{HU20131136} reports an SFE of 0.06~J~m$^{-2}$ using an ANNNI model, which corresponds to 60~mJ~m$^{-2}$ and is noticeably higher than values obtained from DFT-based supercell approaches. This discrepancy is likely attributable to the inherent approximations of the ANNNI model rather than a fundamental physical difference.

Focusing on DFT-based supercell calculations using either rigid shift or tilted cell methods, the basal I2 SFE of Mg consistently falls within the range of approximately 32--36~mJ~m$^{-2}$. Tilted-cell calculations (e.g., Refs.~\cite{wen2009systematic,yin2017comprehensive}) generally yield values near the lower end of this range, while rigid shift approaches cluster around 35~mJ~m$^{-2}$. Overall, these results suggest a robust basal I2 SFE for Mg of about $34 \pm 2$~mJ~m$^{-2}$.

\begin{table}[htbp]
\centering
\caption{I2 Stacking fault energies for Mg from different studies.}
\label{tab:sfe_mg}
\begin{tabular}{cccccccccccc}
\hline
Paper &
k-point mesh &
Cutoff &
Unit &
SFE &
SFE unit &
SF Plane &
Label &
Direction &
SF Type &
Method \\
\hline
Ref.~\cite{DOU2015405} & 7$\times$8$\times$3 & 300 & eV & 36   & mJ m$^{-2}$ & \{0001\} &        & $\langle 1\bar{1}00\rangle$ & Intrinsic & Rigid shift \\
Ref.~\cite{DOU2015405} & 7$\times$8$\times$3 & 300 & eV & 35   & mJ m$^{-2}$ & \{0001\} &        & $\langle 1\bar{1}00\rangle$ & Intrinsic & Rigid shift \\
Ref.~\cite{DOU2015405} & 7$\times$8$\times$3 & 300 & eV & 35   & mJ m$^{-2}$ & \{0001\} &        & $\langle 1\bar{1}00\rangle$ & Intrinsic & Rigid shift \\
Ref.~\cite{DOU2015405} & 7$\times$8$\times$3 & 300 & eV & 35   & mJ m$^{-2}$ & \{0001\} &        & $\langle 1\bar{1}00\rangle$ & Intrinsic & Rigid shift \\
Ref.~\cite{DOU2015405} & 7$\times$8$\times$3 & 300 & eV & 35   & mJ m$^{-2}$ & \{0001\} &        & $\langle 1\bar{1}00\rangle$ & Intrinsic & Rigid shift \\
Ref.~\cite{wen2009systematic} & 15$\times$15$\times$3 & 350 & eV & 34 & mJ m$^{-2}$ & \{0001\} &        & $\langle 1\bar{1}00\rangle$ & I2 & Tilted cell \\
Ref.~\cite{ctx5177087330006471} & 7$\times$7$\times$1 & 400 & eV & 32.5 & mJ m$^{-2}$ &          & Basal  &                             & I2 & Rigid shift \\
Ref.~\cite{HU20131136} & 12$\times$12$\times$6 & 16 & Ry & 0.06 & J m$^{-2}$  &          &        &                             & I2 & ANNNI \\
Ref.~\cite{HU20131136} & 12$\times$12$\times$6 & 16 & Ry & 0.06 & J m$^{-2}$  &          &        &                             & I2 & ANNNI \\
Ref.~\cite{yin2017comprehensive} &  &  &  & 34 & mJ m$^{-2}$ &          & Basal  &                             & I2 & Tilted cell \\
\hline
\end{tabular}
\end{table}

\section{Discussion}
\label{sec:discussion}

Generally, the procedure we propose represents a workaround w.r.t. the lack of standardization of materials data, here specifically computational workflows.
There are essentially two ways to deal with standardization~\cite{Ghiringhelli2017}.
Either one accepts the heterogeneity and data is converted and integrated once it is used or the community develops and agrees on standards which means data is converted and integrated at creation.
Historically, the format of a scientific publication is de facto the standard format for dissemination, i.e. heterogeneity has to be accepted if one is interested in the information presented therein.
At the scale of more than 100's of thousands of papers per year alone in materials science~\cite{Schilling-Wilhelmi2025}, extracting valuable data is not possible by hand.
Consequently, procedures for \textit{post publication research data management} are needed.

We present a specific workflow for the extraction of computational workflows as an example.
However, we expect that our approach from dataset creation based on a specific question (here stacking fault energies in Mg and its alloys), information retrieval (which kind of information is contained in a specific paper), information extraction (unstructured text and tabular data to JSON), and ontology alignment (based on extension of existing ontologies) is general and applicable to other information extraction tasks.
Our various filtering steps and narrowing of scope, both thematically and w.r.t. tokens optimize the final extraction query to the LLM for maximal performance for the specific task.
In short, our approach aims to let the LLM guide itself by providing template-like queries for which the specifics, like the material or method are a user choice. This helps the LLM to deal with the very heterogeneous format that `a scientific paper' is.

In the following, we discuss the individual parts of our extraction workflow and critically discuss the issues we encountered in the hope that this is useful for others.

\subsection{Dataset creation}
The Scopus search alone provides an insufficient filtering mechanism, as seven retrieved records lack abstracts, and there are records either contain irrelevant terms, mention SFE without performing its calculation, or use SFE solely as an input parameter for another model such as ~\cite{LIU201597, FAN20141}.
Furthermore, sometimes information regarding the SFE naturally often appears only in the body of an article, such as, in the last paragraph of the introduction or conclusion, not in title or abstract. For example in Ref.~\cite{DONG20181773}, elements are mentioned in introduction.
Within the automated filtering process, the extracted evidence consistently organizes into several recurring categories as shown in Figure~\ref{fig:sankey_pipeline}:

\begin{enumerate}
    \item Across computational studies, the information associated with the computational method field reflects several distinct methodological approaches.  
    Many papers explicitly describe the use of density functional theory (DFT) or synonymous terminology such as first-principles methods, \emph{ab initio} calculations, plane-wave DFT, generalized gradient approximation (GGA), or spin-polarized GGA. These descriptions appear in varied forms, including statements that calculations were carried out using first-principles approaches, investigated within DFT, or computed using \emph{ab initio} methods, as well as through references to specific implementations such as the Vienna \emph{Ab Initio} Simulation Package (VASP), linearized augmented plane wave methods (LAPW), orbital-free density functional theory, or Kohn--Sham density functional theory.  
    There are also studies in which DFT is combined with additional theoretical or computational frameworks, such as Peierls--Nabarro theory, cluster expansion approaches, or machine-learning interatomic potentials trained on DFT data.  
    Other papers describe molecular dynamics or molecular statics simulations using embedded-atom or modified embedded-atom method interatomic potentials, often under the broader description of atomistic simulations.  
    In addition, some abstracts refer to computational analysis without specifying a clearly identifiable computational method.

    \item Within studies based on DFT, the material systems investigated include pure elemental materials, alloy systems, and combinations of both.  
    Some papers focus on elemental systems, most commonly magnesium, as well as other HCP metals such as titanium, zirconium, rhenium, and cadmium.  
    Other studies examine alloyed systems, including Mg based alloys and non Mg based alloys. Mg based alloy studies include both binary systems and more compositionally complex alloys, such as ternary or higher-order compositions.  
    There are also papers that focus exclusively on alloy systems, as well as studies that analyze both pure metals and their alloyed counterparts for comparison.  
    In some cases such as Ref, abstracts refer generically to ``Mg alloys'' without specifying the number or identity of alloying elements, or describe materials in broad terms such as ``hcp metals'' or ``hexagonal close-packed systems.''

    \item Within the 71 only DFT-based papers, the material systems studied fall into three distinct categories: pure metals, alloys, and studies that treat both pure and alloyed materials.
    A substantial number of works examine only elemental systems, most commonly pure Mg or/and other hcp metals such as Ti, Zr, Re, and Cd where no alloying additions are present.
    Within the alloy and both categories, the systems further divide into Mg-based alloys and non-Mg-based alloys.
    Among the Mg-based studies, many papers explicitly examine binary Mg-X alloys, while others investigate more compositionally complex systems such as ternary Mg-X-Y alloys.
    A second major group consists solely of alloys, the majority of which are Mg-based.
    These range from simple binary Mg-X alloys with a single solute element (Mg-Zn, Mg-Al, Mg-Li, Mg-Sn, Mg-Ca, Mg-Ag, and Mg-RE systems) to more compositionally complex ternary Mg alloys, including Mg-Zn-Y, Mg-Al-Zn, Mg-Sn-X, Mg-Li-Ca, and various Mg-RE-X LPSO systems.
    A smaller subset of alloy-only studies investigate non-Mg-based systems such as Al-Mg, Al-Sc, Al-X, Cu-X, Zn-X, and Mg-doped GaN such as Ref.~\cite{ZHAO2017841, CHISHOLM2001432}.
    In addition, there are papers which fall into a ``not mentioned or implied'' category, where the abstract either provides no explicit description of alloying such as ``Mg alloys'' or ``43 alloying elements'' or refers to generic crystalline systems such as ``hcp metals'' or ``hexagonal close-packed systems'' such as Ref.~\cite{DONG20181773}.
    
    \item Along all DFT-based Mg and Mg-based binary alloy papers, the majority of studies explicitly or implicitly examine materials in the HCP structure, either by directly naming ``hcp metals'', ``hcp systems'', or through discussion of stacking faults and slip behavior characteristic of hcp lattices.
    Many abstracts focus solely on systems described simply as ``hexagonal-close-packed'', ``hcp magnesium'', or ``hcp metals'' without referencing any transformation or alternative crystal phase.
    A smaller subset (four papers) investigates materials in non-hcp crystal structures, including bcc systems (bcc phases stabilized by composition changes)~\cite{SHIN2014198}, B2-structured compounds~\cite{wu2011generalized}, long period stacking order (LPSO)~\cite{PhysRevB.86.054105}, and intermetallic phases that have their own distinct crystal structures (Laves phases)~\cite{ma2013ab}. Mostly the structure is not explicitly mentioned but implied that if the alloy is Mg-based then it has HCP structure.
    
    \item Across DFT-based Mg and Mg-based binary alloys with HCP structure studies, stacking-fault-related energies form the central computational quantity, and the majority of additional properties reported in the abstracts are either directly derived from these fault energies or are obtained through models that use GSFE as an essential input.
    In many papers, no other calculated properties are present beyond SFE itself, reflecting a singular focus on generalized stacking fault energies.
    In the largest group, the additional quantities such as twinnability, Rice criterion, ideal shear strength, activation probabilities of slip systems, and restoring forces are direct functions of the calculated SFE/GSFE, meaning they do not represent independent simulations but rather transformations or interpretations of the fault-energy landscape properties.
    A second class of work reports properties that are directly related to SFE through higher-level modeling frameworks, most commonly the Peierls–Nabarro model, which uses GSFE as a parameter to compute dislocation core widths, Peierls stresses, dissociation distances, or dislocation structure energetics, or machine learning potential development on DFT data; these quantities are semi-direct functions of SFE alone and involve additional theoretical machinery. However, as they add another method to the computational method, they will be excluded by the computational method filter.
    A smaller subset of papers (8) compute properties independent of SFE, such as surface energies, electronic structure, or bonding descriptors, though these are typically used to interpret how alloying influences SFE rather than serving as separate property of interest.

\end{enumerate}

In this study, manual filtering with respect to the parameters affecting SFE calculation is required to keep the analysis tractable and focused.
The main reason is because we focus on the methodological development and, therefore, need a controlled environment with defined inputs and outputs we can match to an expected reference output.
Rather than attempting to capture all possible physical dependencies of the SFE, we explicitly constrain the problem to correlations between computational inputs and their corresponding outputs.
Without such constraints, the parameter space associated with SFE calculations is very heterogeneous, with multiple independent variants for creating a canonical schema. These include variations in slip systems, temperature and stress conditions, alloying effects (such as solute concentration, solute position relative to the stacking fault plane, and occupation sites), structural relaxation protocols, supercell size, and the choice of SFE modeling approach (ANNNI (Ising) model, climbing-image nudged elastic band (CINEB), tilted-cell method, or rigid-shift method). While all of these factors influence SFE, they are not uniformly or consistently reported across the literature.

By manually filtering the corpus to retain only studies that provide sufficiently complete and comparable descriptions of computational workflows and their resulting SFE values, we reduce the effective dimensionality of the problem to a set of well-defined input-output relationships. These workflow-level dependencies, linking methodological choices to calculated SFE values, form the foundation for constructing a minimal, unified, and ontology-aligned schema for the SFE.
This dependency can be written as

\begin{equation}
\gamma_{\mathrm{SFE}} =
f\!\left(
\mathbf{s},
\tau,
T,
\boldsymbol{\sigma},
\alpha,
\mathbf{x}_{\mathrm{solute}},
p_{\mathrm{solute}},
\mathbf{L},
\mathcal{M},
\mathcal{R}
\right),
\label{eq:sfe_dependencies}
\end{equation}

where $\mathbf{s}$ denotes the slip system, $\tau$ the fault type, $T$ the temperature, and $\boldsymbol{\sigma}$ the applied stress. The solute-related parameters include the solute species $\alpha$, its position relative to the stacking fault plane $\mathbf{p}_{\mathrm{solute}}$, and the solute concentration $x_{\mathrm{solute}}$. The supercell dimensions are represented by $\mathbf{L}$, while $\mathcal{M}$ and $\mathcal{R}$ denote the SFE modeling approach and the structural relaxation protocol, respectively.

This schema can be systematically extended or generalized to support the creation of a literature-based canonical schema.
Our overall approach demonstrates that LLM-assisted scientific information extraction becomes reliable only when the conceptual scope of the problem is carefully constrained in particular if a property exhibits complex dependencies or correlations.
Without such constraints, the extraction task quickly becomes inconsistent and, therefore, useless.

In preliminary experiments using a broader and less filtered set of papers, the resulting outputs exhibited substantial schema fragmentation, with semantically similar concepts represented under divergent or incompatible keys, making canonicalization significantly more difficult, i.e. practically impossible.
A concrete example of this effect arises when papers presenting studies that explicitly model temperature-dependent SFEs are included alongside 0\,K calculations.
In this case, additional concepts such as temperature-dependent free energy terms, thermal averaging procedures, or entropy contributions are introduced into the extracted JSON representations.
As a result, the structure of the canonical schema expands to accommodate these new dimensions, increasing both the number of concepts and the complexity of their relationships. This heterogeneity complicates schema unification and reduces comparability across data points.

Moreover, attempting to process all papers simultaneously without prior filtering is impractical from an LLM perspective.
First, the limited context window of current LLMs~\cite{Hsieh2024,Levy2024,Li2024a} prevents the simultaneous consideration of long and heterogeneous methodological descriptions.
Second, when excessive and weakly related information is provided, the resulting knowledge graph representations become sparser: fewer high-quality triplets are extracted, and critical workflow-relevant relationships are often diluted or ignored in favor of more generic associations~\cite{Hsieh2024,Levy2024,Li2024a}.
Our multi-layer filtering strategy mitigates these issues by restricting the analysis to a well-defined subset of papers that share comparable computational workflows and reporting practices. A curated dataset therefore provides a stable foundation for developing consistent, accurate, and interpretable metadata extraction pipelines specifically tailored to SFE calculations in Mg and Mg-based alloys.

\subsection{Information retrieval}
As shown in Figure~\ref{fig:section_cat}, 75.5\% of the papers in the corpus follow the IMRD (Introduction-Methods-Results-Discussion) writing structure. This indicates that keyword-based retrieval is sufficient for a large portion of the dataset, as clearly defined method sections can be reliably identified using section headers and simple keyword matching. Consequently, applying dense retrieval uniformly across all papers is neither necessary nor computationally efficient. In particular, because these papers explicitly contain method sections by construction, keyword matching provides a robust and precise mechanism for retrieving relevant methodological content, whereas dense retrieval requires the generation and storage of embeddings, incurring additional computational and storage overhead.

However, we demonstrate that for papers that do not adhere to standard sectioning conventions,  the enhanced dense retrieval strategy is a viable alternative.
It enables the identification of relevant methodological content without requiring prior knowledge of the paper’s structure or the specific terminology used and requires no user input.
A key challenge in this setting is the selection of the top-$k$ most similar text chunks. Based on empirical evaluation, values of $k$ between 2-5 provide a reasonable trade-off between recall and noise, depending on whether the paper is without sections or has many sections. 

Given the relatively small number of papers within the narrow scope of this study, a statistically rigorous optimization of the top-$k$ parameter is left for future work. In the present work, the value of $k$ is selected heuristically.

Text chunking is performed at the paragraph level to align with the structure of the Markdown representation and to avoid overlapping chunks. However, this strategy has a potential limitation: paragraphs may be split across page boundaries, leading to fragmented content that can negatively affect similarity ranking and top-$k$ selection.

Overall, retrieving method sections from papers with multiple clearly labeled sections is generally more reliable than from papers lacking explicit sectioning, as section headers provide strong semantic anchors that improve retrieval performance.

\subsection{Information extraction}
In terms of the extraction, and specifically for our method extraction prompt, we encountered different problems worthy to note. Our Extraction Prompt~1 states:

\begin{quote}
``You will encounter various data types in these texts, such as chemical compositions, computational\_method, supercell\_details, and fault\_details, often accompanied by specific measurement conditions. Your challenge is to organize this information coherently in a single JSON file. In your approach, pay special attention to the `compositions`. Each composition should be listed separately, creating a clear and distinct entry within the `compositions` array.
Under each composition, you need to provide detailed information in subsections named `computational\_method`, `supercell\_details` and `fault\_details`.'' 
\end{quote}

Based on this instruction, our trials show that a LLM only creates different entries for each composition, while in principle we could also generate different entries for the other subsections (computational method, supercell details and fault details).
For example, we may have different $k$-point meshes depending on different supercell sizes, and we may have different supercell sizes because of different slip systems. 
This means that to represent these dependencies correctly in a serializable JSON, this information would need to be flattened according to these relationships which a LLM is not instructed to do.

However, flattening the hierarchy of a JSON in this way is not straightforward, even for a human.
This is because many of these entries are dependent on each other.
For instance, if we have two different relaxation directions, this could in theory be interpreted as two separate entries, but if the relaxation procedure actually uses both directions together, then we cannot separate them but they must appear together.

For compositions, this problem does not exist, since each composition is truly independent.
There are interdependent parameters and separable parameters such as the dependency between supercell size and $k$-point mesh. The dependency is based on the type of properties, meaning that if they are from same type, there should be no separation. Therefore, we keep the method JSON as it is, without changing its original structure.
It is also worthy to note that we only give the text modality of the method section to the LLM for extraction.

However, in many papers (at least 30 papers), essential methodological details appear in figures or tables within the method section, and connecting them would be necessary for a complete extraction of the presented workflow.
Another challenge originates from our strategy to create a JSON for each method section.
Because subsequently we use the keys contained therein for table extraction.
We observed the following four cases: 

\begin{enumerate}
    \item Parameters that appear only in the method section
    \item Parameters that appear only in the table (such as SFE value and unit, or relaxation procedure)
    \item Parameters that appear in both the table and the method section and have the same representation (such as slip system)
    \item Parameters that appear in both the table and the method section but do not share the same semantics
\end{enumerate}

The most problematic parameters are those with semantic ambiguity; consider the following examples:
\begin{itemize}
    \item If we have ``GGA'' in the method section, it refers to the exchange-correlation functional, but if ``GGA'' appears in the table, the LLM may misclassify it as a computational method~\cite{wen2009systematic}.
    \item If ``L1'' and ``L2'' appear in the method section, they refer to different locations of the stacking-fault plane, but if they appear standalone in a table, the LLM may interpret them as labels rather than physical positions~\cite{WANG2013445}.
\end{itemize}

To solve this issue, we use the keys of the method section JSON during table extraction, while allowing the LLM to add keys when necessary.
This ensures that different modalities (text and tables) remain connected, since treating any modality in isolation will not yield a complete workflow extraction as each modality presents something new while it has many things in common with other modalities.
One may ask why we did not give both the method section and the tables together from the beginning.
The reason is that in our tests, a LLM cannot capture all details when the input is too long (cf.~\cite{Hsieh2024,Levy2024,Li2024a}), it ignores some details, and we lose completeness in the JSON output.
On the other hand, the LLM may generate keys such as ``purpose'' or ``notes'', which are not useful for workflow modeling but can be refined later. 
Refinement is preferable to missing important details.
In the extraction from tables, we consider constraints in the prompt so that we only focus on SFE and its related parameters in the present work:

\begin{enumerate}
    \item If there are reference values from experiments or previous work, we ignore them.
    \item Columns or rows that are not related to the SFE are ignored.
\end{enumerate}

Given that our JSON outputs contain four categories (composition, supercell details, computational method, and fault details), we do not group keys within each of these categories.
The reason is that certain parameters may appear in more than one category during JSON creation, resulting in repetition.
To avoid duplication, we extract all unique key-value pairs and group them in a second step.
A simple example is the parameter \texttt{k\_points}.
For example in Ref~\cite{wen2009systematic}, \texttt{k\_points} belongs to the computational method category, but when different supercell sizes are used, each supercell may have its own \texttt{k\_points}, which places it under supercell details.
This results in a subsequent refining step.

With the canonical schema, and by using both the method section and the table, we can prompt the LLM to create one unified set of method key-value pairs and then treat each SFE value and its associated properties as one data point.
Because we know exactly what we want to extract through the schema, no information is lost.
A remaining issue arises when two keys can represent the same underlying concept, such as \texttt{fault\_name} and \texttt{slip\_plane}.
For example, if the value ``basal stacking fault'' appears, the LLM might map it as \texttt{slip\_plane: basal} or \texttt{fault\_name: basal stacking fault} 
This introduces mixed representations.
We resolve this through ontology-based mapping, because an ontology allows us to detect invalid or inconsistent key-value pairs. 
For the information that is not mentioned, if possible, we tried to have a predefined value. For example, if the composition is not specified, we consider it as pure Mg.

\subsection{Ontology alignment}
During the mapping process, we are not simply mapping key-value pairs to ontological concepts; rather, we map (possibly) non-ontological JSON keys that contain zero conceptual dependencies to an ontology where a calculated property such as the SFE is connected to a sample and is derived from a simulation using a computational method. Table~\ref{tab:mapping_json} shows how different key-value pairs are mapped to ontology-based concepts manually.

\begin{table}[t]
\centering
\caption{Mapping between extracted JSON fields and ontology-based JSON representation.}
\label{tab:mapping_json}
\small
\setlength{\tabcolsep}{5pt}
\renewcommand{\arraystretch}{1.2}
\begin{tabular}{lp{6.5cm}p{6.5cm}}
\hline
Category & Extracted JSON Field & Ontology JSON Mapping \\
\hline

\multirow{9}{*}{\parbox{3.8cm}{Computational method mapping}}
& \texttt{"dft\_software": "VASP"}
& \texttt{"SoftwareAgent": "VASP"} \\

& \texttt{"xc\_functional": "GGA-PW91"}
& \texttt{"XC\_EnergyFunctional": "GGA-PW91"} \\

& \texttt{"pseudopotential": "PAW"}
& \texttt{"InputParameter": \{"Pseudopotential": \{"label": "PAW"\}\}} \\

& \texttt{"plane\_wave\_cutoff\_energy": "350 eV"}
& \texttt{"EnergyCutoff": \{"value": 350.0, "unit": "EV"\}} \\

& \texttt{"energy\_convergence\_criterion": "1e-6 eV/atom"}
& \texttt{"EnergyConvergence": \{"value": 1e-6, "unit": "EV"\}} \\

& \texttt{"force\_convergence\_criterion": "1e-2 eV/A"}
& \texttt{"ForceConvergence": \{"value": 0.01, "unit": "EV-PER-ANGSTROM"\}} \\

& \texttt{"k\_point\_grid": "15x15x3"}
& \texttt{"KpointMesh": \{"value": [15, 15, 3]\}} \\

& \texttt{"k\_point\_grid\_type": "Monkhorst-Pack"}
& \texttt{"KpointMesh.Type": "MonkhorstPackKPointMesh"} \\

& \texttt{"calculation\_method": "supercell tilting technique"}
& \texttt{"Simulation": "Tilted Cell"} \\
\hline

\multirow{4}{*}{\parbox{3.8cm}{Supercell to computational sample mapping}}
& \texttt{"supercell\_size": "1x1x6"}
& \texttt{"hasRepetition\_x": 1, "hasRepetition\_y": 1, "hasRepetition\_z": 6} \\

& \texttt{"number\_of\_atoms": "12"} + \texttt{"composition": "Mg"}
& \texttt{"ChemicalComposition": "Mg12"} \\

& \texttt{"slip\_plane": "\{0001\}"}
& \texttt{"hasSFplane": "(0001)"} \\

& (Inferred) Mg $\rightarrow$ hcp structure
& \texttt{"CrystalStructure": "hcp"} \\
\hline

\multirow{3}{*}{\parbox{3.8cm}{Fault metadata mapping}}
& \texttt{"fault\_label": "I1"}
& \texttt{"Type": "I1StackingFault"} \\

& \texttt{"fault\_type": "basal stacking fault"}
& \texttt{"SF\_hcp\_label": "basal"} \\

& \texttt{"slip\_plane": "\{0001\}"}
& \texttt{"hasSFplane": "(0001)"} \\
\hline

\multirow{3}{*}{\parbox{3.8cm}{SFE mapping}}
& \texttt{"stacking\_fault\_energy": 16}
& \texttt{"value": 16} \\

& \texttt{"stacking\_fault\_energy\_unit": "mJ/m\^2"}
& \texttt{"unit": "http://qudt.org/vocab/unit/MilliJ-PER-M2"} \\

& \texttt{"fault\_name": "I1 stacking fault"}
& \texttt{"label": "stable"} \\
\hline

\end{tabular}
\end{table}

For the rigid-shift and tilted-cell approaches, we need to define two computational samples, because the SFE is the energy difference between a perfect structure and the stacking-fault structure.
For the ANNNI-model papers, we have one computational sample for each crystal structure, cf.~\ref{sssec:sfe_workflows}.

Once the ontology-based JSON is created, we use it as input to the atomRDF SFE workflow so that the SFE value can, in principle, be reproduced. However, this step is neither easy nor straightforward, for two major reasons:
\begin{itemize}
    \item The input JSON does not contain all the required information\\
    Several essential metadata fields are often missing, incomplete, or only partially available:
    \begin{itemize}
        \item \textbf{SimulationCell metadata} is not fully described in the text in in all of our eight papers.
        \begin{itemize}
            \item \texttt{SimulationCellVector} is often shown only in a figure illustrating the supercell.
            \item \texttt{SimulationCellLength} and \texttt{SimulationCellAngle} may appear only in a figure caption, if at all mention it.
            \item \texttt{hasVacuumLength} is sometimes mentioned in the method section, but may also appear only in a figure caption.
            \item \texttt{hasVacuumDirection} is almost never explicitly stated, it is assumed in all the papers (``10~\AA{} vacuum normal to the slip plane'').
        \end{itemize}
    
        \item \textbf{StackingFault metadata} is inconsistently described.
        \begin{itemize}
            \item Authors may state only the fault type (``I2'', ``E'', ``T2'') without explicit mentioning of slip plane.
            \item Sometimes only the family of planes is mentioned without Miller indices.
            \item \texttt{hasDisplacementVector} may include a magnitude but no direction, or vice versa.
            \item \texttt{hasRelativeDistance} (distance between solute and SF plane) is almost never exactly reported in cartesian coordinate.
        \end{itemize}
    \end{itemize}

These last issues create further ambiguity and can only be guessed by experiential knowledge:

    \begin{itemize}
        \item The exact location such as Cartesian coordinate of the SF plane is never stated.
        \begin{itemize}
            \item Rigid-shift calculations $\rightarrow$ usually the middle layer.
            \item Tilted-cell calculations $\rightarrow$ one SF at the top, one at the bottom.
            \item ANNNI $\rightarrow$ no explicit SF plane.
        \end{itemize}
        \item Solute placement (``on the SF plane'', ``near the SF plane'') lacks precise coordinates.
        \item Even if a layer number is stated (``layer~4''), authors seldom define the origin or counting direction~\cite{dou2019generalized}.
        \item Even with layer information, when a layer contains multiple atoms, the specific substitution site is unknown.
    \end{itemize}

Such metadata either appears only in figures (requiring interpretation) or is absent altogether. 
In the \texttt{ComputationalMethod} group, the \texttt{Simulation} type (rigid shift, tilted cell) is often implied rather than explicitly stated. 
For example, the phrase ``upper half was shifted relative to lower half'' implies a rigid shift.
Relaxation procedures are often stated vaguely or not at all.
    
Stacking-fault values in tables may come from a different simulation whose metadata is not described in the method section.
Some tables list SFE values without specifying the stacking-fault type, the slip plane or the slip system requiring cross-referencing with the results text.
Some compositions listed in the method section do not appear in the SFE table; instead, their values may appear in figures or purely in the text.

\item Several workflow steps are not described in the paper
    There are steps in the actual computational workflow that researchers almost always
    For example:
    \begin{itemize}
        \item Structural minimization or relaxation prior to fault creation
        \item Details of partial relaxations or constrained relaxations
    \end{itemize}
\end{itemize}

All these values are essential for reproducing SFE values, yet they are typically not explicitly documented in the method section of papers.

Even after constructing a canonical schema that enables structural alignment and direct comparison across studies, meaningful comparison of results is not guaranteed without additional post-processing. The numerical values themselves may still be expressed in incompatible formats or units, requiring further normalization before they can be compared.

A common example is the unit of stacking fault energy, which may be reported in J\,m$^{-2}$, mJ\,m$^{-2}$, or other equivalent units. Without an explicit unit-conversion step, these values are not directly comparable, despite being stored under the same canonical field.

Beyond numerical normalization, semantic normalization is also required. If the schema faithfully preserves the terminology used in the source text, the same physical defect may be labeled inconsistently across papers. For instance, some studies explicitly distinguish between I$_1$ and I$_2$ intrinsic stacking faults, while others refer only to an ``intrinsic stacking fault'' without further specification. Treating these labels verbatim results in ambiguities that hinder comparison, aggregation, and downstream analysis. Resolving such cases requires additional post-processing to map textual descriptions onto a controlled vocabulary or defect taxonomy.

Summarizing the discussion, our work presents a rigorous, ontology-aligned workflow for the automated extraction and reuse of computational workflows from scientific literature, specifically for SFE calculations in magnesium and its alloys.
The core insight is that successful LLM-based extraction demands not just sophisticated prompting, but a very strict, multi-layered filtering strategy to constrain the problem space. Without such filtering, the inherent heterogeneity of scientific writing (varying terminology, incomplete reporting, and inconsistent structural descriptions) leads to schema fragmentation, making canonicalization of keys across papers impossible.
By systematically narrowing the corpus to papers with comparable computational methodologies (DFT, HCP structure, binary Mg-based alloys, explicit SFE focus), we create a stable, tractable environment where LLMs can reliably extract and align structured data.
This filtering is not merely a preprocessing step.
It is a methodological necessity that enables the entire pipeline.
The LLM is guided by a carefully engineered prompt architecture that uses both keyword-based and in rarer cases dense retrieval to locate relevant methodological and tabular content.

The extraction process is not a one-shot operation. It involves a three-stage workflow: (1) initial extraction of methodological details and table data into heterogeneous JSONs, (2) semantic grouping and unification into a canonical schema using LLMs with moderate temperature ($T=0.3$), and (3) final mapping of this schema to a formal ontology (CMSO~\cite{CMSO}, ASMO~\cite{ASMO}, PLDO~\cite{PLDO}, and their extensions) to create a machine-readable knowledge graph.
Such a multi-stage approach with its guidance rooted in domain knowledge and iterative refinement is essential for overcoming the LLM's tendency to hallucinate or misinterpret context when faced with unstructured, multimodal data.

The resulting ontology-aligned knowledge graph within the atomRDF framework enables downstream applications such as querying for SFE values across different slip systems, comparing results from different modeling approaches (rigid-shift vs. tilted-cell vs. ANNNI).
Automated reproduction of published workflows, however, remains a challenge mainly because of missing parameters that can only be guessed based on experiental knowledge.
In other words, the semantic richness of the extracted data often falls short of the computational completeness required for true reproduction.
Missing metadata, such as simulation cell vectors, vacuum directions, exact solute positions, and relaxation protocols, means that even with a perfect knowledge graph, exact reproduction remains practically impossible without additional user input.

This underscores a key perspective for applying such a workflow to experimental papers: while computational workflows are inherently procedural and thus, in principle, amenable to formalization, experimental workflows involve manual, context-dependent steps that are far harder to capture \textit{automatically}.
Nevertheless, the principles of our approach are in principle directly transferable.
For experimental papers, the focus would shift to capturing procedural details (e.g., sample preparation, measurement conditions, calibration steps) with the same level of rigor, using ontologies for experimental methods and materials. The ultimate goal remains the same: to transform the vast, unstructured literature into a FAIR, reusable, and reproducible knowledge base, or short: post-publication research data management.

\section{Conclusions}
\label{sec:conclusion}

Our work establishes a robust, ontology-driven framework for the automated extraction, structuring, and reuse of computational workflows from scientific literature, demonstrating its general viability for reproducing stacking fault energy calculations in magnesium and its alloys.
By combining a stringent, multi-stage filtering strategy with carefully engineered prompt engineering and a formal ontology alignment process, we overcome the inherent challenges of heterogeneity and ambiguity in scientific texts.

The key innovation lies in the strict guidance of the LLMs through a workflow that first narrows the scope to a homogeneous, methodologically comparable subset of papers, then systematically extracts and unifies information from text and tables into a canonical schema, and finally maps this schema to established materials science ontologies (CMSO, ASMO, PLDO~\cite{CMSO, ASMO,PLDO}). Our approach ensures that the extracted data is not only structured but also semantically interoperable, enabling the creation of a knowledge graph that supports advanced querying and reasoning.

The resulting knowledge graph provides a foundation for three applications: (1) systematic comparison of SFE values across different materials and slip systems, (2) the still challenging, but theoretically possible automatic reproduction of published results by translating the ontology-aligned metadata into executable workflows within tools like atomRDF if a human supplies experiential knowledge, and (3) the reuse of validated simulation protocols for new parameter studies.
While the current implementation reveals that complete computational reproducibility is still hampered by missing or implicit metadata, the framework provides a clear path forward.

Future work will focus on improving the extraction pipeline to infer missing parameters from figures and contextual clues (cf.~\cite{Wang2025a} for an agentic approach to create computational workflows in DFT based on natural language input or~\cite{Zimmermann2025a,LangSim} for a Large Language Model interface for atomistic simulations), and on the development of validation mechanisms. 
As an example, frameworks such as LangSim~\cite{LangSim}, an LLM extension that provides agent-based interfaces for coupling large language models with scientific simulation codes, can directly use the ontology-based JSON representation to run simulations and compute physical properties from natural language inputs, without requiring further post-processing of the JSON data.

Beyond our specific use case, the methodology presented here offers a generalizable approach for the digital transformation of materials science literature. It demonstrates that LLMs, when guided by domain-specific constraints and integrated with formal ontologies, can become useful tools for extracting and structuring complex scientific knowledge.
We hope that our procedure is useful for the community, and hope that the community moves towards a future where computational and experimental workflows are reported in standardized, machine-readable formats from the outset, thereby ensuring the long-term reproducibility, reusability, and trustworthiness of scientific research output.

\medskip
\textbf{Acknowledgements} \par 
MS gratefully acknowledges funding by Deutsche Forschungsgemeinschaft (DFG) for CRC1625, project number 506711657, subprojects A05, INF, and financial support by the European Union by ERC Grant DISCO-DATA, Project No. 101161287. The views and opinions expressed are, however, those of the authors only and do not necessarily reflect those of the European Union or the European Research Council Executive Agency. Neither the European Union nor the granting authority can be held responsible for them.

\medskip

%

\printbibliography

\newpage
\appendix
\section{Appendix: Prompts}
\renewcommand\thelstlisting{A.\arabic{lstlisting}}
\setcounter{lstlisting}{0}
\begin{lstlisting}[basicstyle=\ttfamily\small, breaklines=true, frame=single, caption={Filtering Prompt}, label={lst:filtering}]
Study Type Prompt

SYSTEM_PROMPT = """
You are an expert in materials science.
Decide the TYPE OF STUDY based ONLY on the title and abstract.
Title: {title}
Abstract: {abstract}
You must output a JSON object with exactly this structure:

{{
  "Study_type": {{
    "value": "<computational | experimental | both>",
    "evidence": "<short phrase from the text>"
  }}
}}

Rules:
- "computational": main results come from calculations or simulations
- "experimental": main results come from experiments with no clear computational work.
- "both": both experimental AND computational work are clearly present and not one of them used for comparision.

Base your answer ONLY on the given text.
Do NOT output anything except the JSON object.
"""

Computational Method Prompt

SYSTEM_PROMPT = """
Determine whether this study uses density functional theory (DFT) as an atomistic electronic-structure method, based ONLY on the title and abstract.
Title: {title}
Abstract: {abstract}
Output a JSON object with exactly this structure:

{{
  "Comp_method": {{
    "value": "<DFT | Non_DFT | DFT+other | Not_mentioned>",
    "evidence": "<short phrase from the text or an explicit note>"
  }}
}}

Definitions:
- "DFT":
    Use this ONLY if the text clearly establishes that atomistic calculations are performed using first-principles electronic-structure theory, leaving no reasonable interpretation as a classical or parameterized model.

- "Non_DFT":
    Use this ONLY if the study is clearly computational and atomistic, but the primary atomistic method is explicitly not first-principles electronic-structure theory. 
    
- "DFT+other": 
    Use this if the study explicitly combines DFT with other computational methods. 

- "Not_mentioned":
    Use this if the study is classified as computational, but the abstract/title does NOT clearly specify which computational method is used, OR the method is too vague to decide DFT vs non-DFT.

Requirements:
- Always fill BOTH "value" and "evidence".
- "evidence" should be a short phrase or sentence, such as a quote from the abstract or a brief justification (e.g. "no method name appears in the abstract").

Base your answer ONLY on the given text.
Do NOT output anything except the JSON object.
"""


Material Classification Prompt

SYSTEM_PROMPT = """
You are an expert in material classification, with specific expertise in metals and alloys.
Classify the MATERIAL aspects of the study based ONLY on the given title and abstract.
Title: {title}
Abstract: {abstract}
Output a single JSON object with EXACTLY the following structure:

{{
  "Material_type": {{
    "value": "<pure_metal | alloy | alloy+metal |  Not_mentioned >",
    "evidence": "<short phrase from the text or an explicit justification>"
  }},
  "Mg_based_alloy": {{
    "value": "<Yes | No | >",
    "evidence": "<short phrase from the text or an explicit justification>"
  }},
  "Alloy_system_type": {{
    "value": "<binary | non_binary | Not_mentioned>",
    "evidence": "<short phrase from the text or an explicit justification>"
  }}
}}

Here, ">" denotes an intentionally empty string "".

Definitions and rules
1) Material_type
Choose exactly one of the following:
- "pure_metal":
    Use this ONLY if the study clearly investigates elemental metals only, with no alloys mentioned.

- "alloy":
    Use this ONLY if the study clearly investigates alloy systems.

- "alloy+metal":
    Use this if both pure metals and alloys are clearly investigated within the same study.
    
- "Not_mentioned":
    Use this if the abstract/title does NOT clearly specify which material type is used.

2) Conditional behavior

A) If "Material_type" == "pure_metal":
- Do NOT infer any alloy-related information.
- Set the following fields to empty (""):
  - "Mg_based_alloy"
  - "Alloy_system_type"
- In each case, the "evidence" must explicitly state that only pure metals are studied and no alloys are involved.

B) If "Material_type" == "alloy" or "alloy+metal":
You MUST determine whether the alloy system(s) are Mg-based.

2.1) Mg_based_alloy

Choose one:
- "Yes":
    Magnesium is clearly the base element of the alloy system.

- "No":
    The alloys are clearly not Mg-based, or Mg appears only as a minor alloying element in a non-Mg base.

2.2) If "Mg_based_alloy" == "No":

- Do NOT classify the alloy system type.
- Set the following field to empty (""):
  - "Alloy_system_type"
- The "evidence" must clearly state that the alloys are not Mg-based.

2.3) If "Mg_based_alloy" == "Yes":
You MUST classify the Mg-based alloy system type.

Alloy_system_type
Choose one:
- "binary":
    Exactly one alloying element with Mg as the base metal.
    
- "non_binary":
    Two or more alloying elements with Mg as the base metal,OR the text clearly indicates ternary, quaternary, or multicomponent Mg-based alloys.

- "Not_mentioned":
    Use this if the text is too vague to decide, or if alloying elements are not specified at all(e.g., the text only says "Mg alloys" without naming elements).

General requirements
- ALWAYS fill BOTH "value" and "evidence" for ALL four keys.
- For any empty value (""), the "evidence" must explicitly explain WHYthe field is left blank (e.g. pure metal only, not Mg-based, non-binary system).
- If multiple materials or material systems are mentioned, consider each of them individually and then determine the final classification based on the combined set of materials investigated.
- Base your classification STRICTLY on the given title and abstract.
- Do NOT infer, assume, or use external knowledge.
- Do NOT output anything except the JSON object.
"""

Crystal Structure Prompt

SYSTEM_PROMPT = """
You are an expert in crystallography and magnesium-based materials.
Your task is to determine whether the PRIMARY material system investigated in this study adopts a hexagonal close-packed (HCP) crystal structure, based ONLY on the title and abstract.
Title: {title}
Abstract: {abstract}
Output a single JSON object with EXACTLY the following structure:

{{
  "Crystal_structure_HCP": {{
    "value": "<Yes | No>",
    "evidence": "<short phrase from the text or an explicit justification>"
  }}
}}

Classification rules
Choose exactly one of the following:
- "Yes":
    Use this if the title or abstract clearly indicates that the primary system studied has a hexagonal close-packed crystal structure. This includes:
    - explicit mention of HCP or hexagonal close-packed structure, OR
    - clear focus on magnesium or magnesium-based alloys studied, with no indication of a different crystal structure being the main focus.

- "No":
    Use this if the primary system studied is clearly NOT HCP, including cases
    where:
    - a different crystal structure is explicitly stated as the main phase OR
    - the study focuses on Mg-based systems in a clearly non-HCP structural form.

General requirements

- Base your decision STRICTLY on the given title and abstract.
- Do NOT infer unstated crystal structures.
- If multiple materials are mentioned, base the decision on the PRIMARY system investigated.
- Always provide a clear "evidence" justification.
- Do NOT output anything except the JSON object.

"""

SFE Relavance Prompt

SYSTEM_PROMPT = """
You are an expert in stacking fault energy (SFE), generalized stacking fault energy (GSFE), generalized planar fault energy (GPFE), and planar fault energetics in pure metals and alloys.
Your task is to assess how strongly this study is focused on the CALCULATION of SFE/GSFE/GPFE/planar fault energies, based ONLY on the title and abstract.
Title: {title}
Abstract: {abstract}
Output a single JSON object with EXACTLY the following structure:

{{
  "SFE_primary_focus": {{
    "value": "<Yes | No >",
    "evidence": "<short phrase from the text or an explicit justification>"
  }},
  "SFE_only_primary_focus": {{
    "value": "<Yes | No | >",
    "evidence": "<short phrase from the text or an explicit justification>"
  }},
  "Other_properties_besides_SFE": {{
    "value": "<Directly_related | Indirectly_related | >",
    "evidence": "<short phrase from the text or an explicit justification>"
  }}
}}

Key interpretation rule

In this task, SFE/GSFE/GPFE/planar fault energies must be EXPLICITLY CALCULATED in the study ( via first-principles electronic-structure theory such as density functional theory).
If these quantities are mentioned only qualitatively, as background, or without a clear indication that they are actually computed, then they must NOT be treated as a primary focus.

Classification rules
1) SFE_primary_focus
Choose one:
- "Yes":
    The study clearly and unambiguously calculates stacking fault energy, generalized stacking fault energy, generalized planar fault energy, or equivalent planar fault energetics, and these
    calculations are a primary or major focus of the paper.

- "No":
    SFE/GSFE/GPFE is absent, mentioned only qualitatively, or appears only as a minor or secondary aspect of the study.

Conditional behavior

If "SFE_primary_focus" == "No":
- Do NOT perform any further SFE-related classification.
- Set the following fields to empty (""):
  - "SFE_is_only_primary_focus"
  - "Other_properties_besides_SFE"
- In each case, the "evidence" must explicitly explain why (e.g. "SFE only mentioned in background" or "no SFE calculation indicated").

2) If and ONLY if "SFE_primary_focus" == "Yes":
You MUST determine whether other major CALCULATED properties are also a focus.
SFE_is_only_primary_focus
Choose one:
- "Yes":
    SFE/GSFE/GPFE is essentially the ONLY major calculated property in the study.

- "No":
    In addition to SFE/GSFE/GPFE, the study clearly calculates other major physical or mechanical properties.

Conditional behavior

If "SFE_is_only_primary_focus" == "Yes":
- Set:
  - "Other_properties_besides_SFE"."value" = ""
- The "evidence" must explain that no other major calculated properties beyond SFE/GSFE are clearly present.

If "SFE_is_only_primary_focus" == "No":
You MUST classify how the additional calculated properties relate to SFE.

Other_properties_besides_SFE
Choose one:
- "Directly_related":
    The additional major properties are primarily direct consequences of, or are explicitly derived from, the computed SFE/GSFE/GPFE, or has an effect on SFE value.

- "Indirectly_related":
    The additional major properties are computed using additional models, simulations, or theoretical frameworks, where SFE/GSFE is only one of several inputs or is not a simple direct function of the SFE surface.

General requirements
- ALWAYS fill BOTH "value" and "evidence" for ALL three keys.
- For any empty value (""), the "evidence" must explicitly state WHY the
  field is left blank.
- Base your classification STRICTLY on the given title and abstract.
- Do NOT infer, assume, or use external knowledge.
- Do NOT output anything except the JSON object.

"""
\end{lstlisting}

\begin{lstlisting}[basicstyle=\ttfamily\small, breaklines=true, frame=single, caption={Extraction Prompt 1}, label={lst:Extraction_prompt_1}]
SYSTEM_PROMPT = """
You are an efficient data-transformation assistant specializing in materials science.
Your task is to process individual excerpts from scientific papers describing fault-energy calculation workflows, translating any markup language into clear text. 
Approach each excerpts as an independent unit, focusing solely on its content without referencing or recalling information from other excerpts. Convert these excerpts into a structured JSON format. 
You will encounter various data types in these texts, such as chemical compositions, computational_method, supercell_details, and fault_details, often accompanied by specific measurements conditions.
Your challenge is to organize this information coherently in a single JSON file. 
In your approach, pay special attention to the 'compositions'. Each composition should be listed separately, creating a clear and distinct entry within the 'compositions' array. 
Under each composition, you need to provide detailed information in subsections named 'computational_method', 'supercell_details' and "fault_details'.
When dealing with numerical data, such as measurements values, remember to include these figures in a 'value' field and specify, their corresponding units in a 'unit' field.

IMPORTANCE: 
1. Reproducibility
    - Extract all the parameters which allows someone to re-run the workflow. 
2. Connection
    - If in the JSON, there are keys which thier values are depending on each others, then make this connection clear.
3. Format 
    - If we have values like this "5 X 5 X 5", keep it as it is. 
"""
\end{lstlisting}

\begin{lstlisting}[basicstyle=\ttfamily\small, breaklines=true, frame=single, caption={Extraction Prompt 2}, label={lst:Extraction_prompt_2}]
SYSTEM_PROMPT = """
As an efficient data transformation assistant specializing in materials science, your task is to process individual tables extracted from materials science papers, translating any markup language into clear text.
Approach each table as an independent unit, focusing solely on its content without referencing or recalling information from other tables. Convert these tables into a structured JSON format.
You will encounter various data types, such as chemical compositions, computational_method, supercell_details, fault_details, often accompanied by specific simulation or modelling conditions. Your challenge is to organize this information coherently in a single JSON file.
In your approach, pay special attention to the 'compositions'. Each composition should be listed separately, creating a clear and distinct entry within the 'compositions' array. Under each composition, you need to provide detailed information in subsections named 'computational_method', 'supercell_details', 'fault_details'.

Precise rules:
1. Schema using 
    - Use the JSON schema keys as hint to create the Table JSON.  
    - If the table introduces a relevant parameter that is NOT present as a key in the schema, ADD a new key for it in the most appropriate section and fill it.

2. Focus only on stacking/planar faults and their energies:
   - Include all the quantities and their data related to value of stacking fault energy (SFE), generalized stacking fault energy (GSFE), planar fault energy (PFE), and generalized planar fault energy (GPFE).
   - Include all the quantities and their data related to stacking fault energy (SFE), generalized stacking fault energy (GSFE), planar fault energy (PFE), and generalized planar fault energy (GPFE).
   - Ignore columns or rows that clearly describe unrelated quantities.

3. Compositions:
   - Each distinct material or alloy must become a separate entry in the "compositions" array.
   - If the table uses a base matrix plus "alloying element" column, combine them into a human-readable composition string such as "base matrix-alloying element" where appropriate.

4. Numerical values and units:
   - Every numeric physical quantity must be represented with a {"value": ..., "unit": ...} object.
   - If the unit is clearly given in the caption or header, use a consistent plain-text representation.
   - Do NOT invent units; if the unit is truly not given, set "unit": null.

5. Referenced vs. "this work" data (very important):
   - Only include values that are explicitly calculated in the present work (often labelled as "this work" or implied by the main body of the table).
   - Exclude any values that are clearly cited from other works:
     - Entries with explicit references such as [2], [21], (DFT)^a, (Exp.)^b, superscripts like ^{a}, ^{b}, ^{c}, or labels that clearly point to other works.
   - If a table cell mixes multiple values and any of them are marked as referenced or as literature/experimental data, exclude the entire cell from extraction.

6. LaTeX and markup cleanup:
   - Convert LaTeX expressions (\{0001\}, \langle 11\bar{2}0\rangle, \gamma_{\rm SF}, \gamma_{\rm USF}) to clean, readable strings.
   - Remove LaTeX wrappers such as \(\) \), \{ \}, \rm, while preserving the underlying meaning (e.g. "{0001}<11-20>" or "gamma_USF").

7. Missing information:
   - If a field is not present for a composition, use an empty array ([]) or null for that field, rather than inventing data.
   - Never guess or infer numerical values.


Output format:

- Return a single valid JSON object.
- It must follow the base schema structure, with filled values where appropriate.
- You must extract all the data points related to the this work.
- You can omit the keys of base schema which has value null/empty. 
- Newly introduced keys from the table must be included in the appropriate section as described above.
"""

\end{lstlisting}

\begin{lstlisting}[basicstyle=\ttfamily\small, breaklines=true, frame=single, caption={Grouping Prompt}, label={lst:Grouping_prompt}]
SYSTEM_PROMPT = """
You are a computational materials science expert specializing in stacking fault energy (SFE) or planar fault energy (PFE) calculations.
Your task is to group the following key-value pairs into a set of conceptual categories based on scientific meaning, not wording.

Grouping Rules
1. A "group" should represent a single physical or computational concept in SFE/PFE modeling.
2. Keys that refer to the same scientific concept must be grouped together even if their names differ, are nested differently, or use inconsistent terminology.
3. Each group must include:
   - a `"group_name"` summarizing the concept
   - a `"key_value_pairs"` object containing all original keys and values belonging to that concept.
4. Do not merge, rewrite, or reinterpret the values. Simply group the original keys.
5. The output must be a valid JSON array, using this schema:

[
  {{
    "group_name": "some_conceptual_category",
    "key_value_pairs": {{
      "key1": [...],
      "key2": [...],
      ...
    }}
  }},
  ...
]
6. Use clear, domain-relevant group names drawn from materials science. 
"""
\end{lstlisting}

\begin{lstlisting}[basicstyle=\ttfamily\small, breaklines=true, frame=single, caption={Canonical Keys Prompt}, label={lst:canonical_keys_prompt}]
SYSTEM_PROMPT = """
You are a data-normalization expert for computational materials science metadata.
Your task is: given a JSON "group" with fields:
- "group_name": string
- "key_value_pairs": object mapping original keys to lists of values
produce a minimal set of "canonical keys" that best describe the group, and map original keys into those canonical keys.

High-level goals:
- Each group should have only a small, meaningful set of canonical keys.
- Canonical keys should be reusable across JSON, not JSON-specific.
- Keys that have the same meaning should be merged.

Rules:
1. Group-level concept  
   - Treat "group_name" as the main topic.
   - Canonical keys should represent natural sub-concepts within this topic.
2. Merge by key-name semantics  
   - Consider dots in keys as hierarchy/qualifiers (Example: "gsfe_calculation.energy_convergence.value" refers to energy convergence).
   - Merge keys that clearly answer the same question:
     - Example 1: "software" and "DFT_code" to one canonical key like "dft_software".
     - Example 2: "exchange_correlation" and "exchange_correlation_functional" to    "xc_functional".
     - Example 3: "vacuum_gap" and ""vacuum_width" to "vacuum_width"
   - Merge nested variants when they clearly describe the same concept:
     - Example: "pseudopotential" and "geometry_optimization.pseudopotential" to "pseudopotential".
3. Canonical naming rules  
   - Short, reusable names in snake_case.
   - Use suffixes "_description", "_note", "_energy", "_scheme".
4. All original keys must appear in at least one canonical "merged_from" list.
5. Return ONLY valid JSON:
{
  "group_name": "...",
  "canonical_keys": [
    {
      "name": "canonical_key_name",
      "description": "short description",
      "merged_from": ["original_key1", "original_key2"]
    }
  ]
}
Return ONLY valid JSON, no extra text.
"""
\end{lstlisting}

\begin{lstlisting}[basicstyle=\ttfamily\small, breaklines=true, frame=single, caption={Extraction Prompt 3}, label={lst:Extraction_prompt_3}]
SYSTEM_PROMPT = """
You are a scientific information extraction system specialized in stacking fault energy (SFE), generalized stacking fault energy (GSFE), and slip system metadata for HCP metals such as Mg and Mg alloys.

Your task:
1. Read a scientific text containing:
   - A "Method" section
   - A table with stacking fault energy values
2. Use ONLY the canonical keys listed below.  
   When filling JSON output:
   - Use values from the text exactly (no guessing).  
   - Leave fields out entirely if not present.  
   - Do NOT invent values.  
   - Detect "this work" vs "literature" using superscripts or notes.

3. Produce two JSON sections:
   - method: information extracted from the method.
   - data_points: one JSON object per row in the table, containing fault energy values and its metadata.

4. The JSON you output MUST use only the canonical keys below.
[Canonical keys]

IMPORTANT RULES:
- Output MUST be valid JSON only.
- If information is missing, omit the key.
- Include all the quantities and their data related to value of stacking fault energy (SFE), generalized stacking fault energy (GSFE), planar fault energy (PFE), and generalized planar fault energy (GPFE).
- Include all the quantities and their data related to stacking fault energy (SFE), generalized stacking fault energy (GSFE), planar fault energy (PFE), and generalized planar fault energy (GPFE).
- Ignore columns or rows that clearly describe unrelated quantities.
- Only include values that are explicitly calculated in the present work (often labelled as "this work" or implied by the main body of the table).
- Exclude any values that are clearly cited from other works: Entries with explicit references such as [2], [21], (DFT)^a, (Exp.)^b, superscripts like ^{a}, ^{b}, ^{c}, or labels that clearly point to other works.
- If a table cell mixes multiple values and any of them are marked as referenced or as literature/experimental data, exclude the entire cell from extraction.
- You must extract all the data points related to the "this work". 
- Do not have any senetnce as value of a key as long as you can make it to multiple keys. 
- Every numeric value of SFE in the table corresponds to EXACTLY one data_points entry. For example: If the row has 8 numeric values, produce 8 JSON objects.
- If your output has fewer objects, it is incorrect.

OUTPUT FORMAT
{
  "method": {
     ... extracted using canonical keys ...
  },
  "data_points": [
     {
        ... one per SFE/USFE measurement ...
     }
  ]
}
"""
\end{lstlisting}

\begin{lstlisting}[basicstyle=\ttfamily\small, breaklines=true, frame=single, caption={Abstract JSON Prompt}, label={lst:Abstract_JSON_prompt}]
SYSTEM_PROMPT = """
You are an expert in computational materials science. Your task is to analyze a scientific abstract and extract specific information in a step-by-step manner, then output a single JSON object with the schema below.
Steps:
1. Read the abstract carefully.
2. Identify the material composition(s)/system(s) studied. Put them in "composition".
3. Identify the computational method(s) and tool(s) used. Put them in "computational_method".
4. Extract any supercell-geometry details 
5. Extract fault/defect-specific details and store them inside "fault_details", including


Output schema (return ONLY this JSON object; no extra text):
{
  "composition": [],
  "computational_method": [],
  "supercell_details": [],
  "fault_details": []
}

- If any of these fields are not mentioned then keep it as empty list. 
- For each key, create list of key-value pairs. 
- do not use reasoning be faithful to the text.
"""
\end{lstlisting}

\begin{lstlisting}[basicstyle=\ttfamily\small, breaklines=true, frame=single, caption={Question Generating Prompt}, label={lst:Question_generating_prompt}]
SYSTEM_PROMPT = """
You are a computational materials scientist reading a scientific paper.
Below is structured information extracted from the abstract of the paper. For each key-value pair, generate a scientific chain-of-thought that:
- Reflects how a domain expert would think about finding this information in the full paper.
- Justifies why and where such information is typically discussed.
- Ends with a clear goal for retrieving relevant chunks of the paper.

Your goal is to create a detailed reasoning prompt (chain of thought) for each pair, which will then be used for retrieving the most relevant sections of the paper.
Output a dictionary where each key maps to a single natural-language chain-of-thought prompt.
"""

\end{lstlisting}

\section{Appendix: Canonical schema prompt temperature dependency}
The following figures show the temperature dependency of the number of canonical keys as heatmaps, with one heatmap for each grouping temperature. The grouping temperature is given in each figure title. All figures follow the same structure: the y-axis lists the group names, and the x-axis shows the second LLM temperature (generating the canonical keys) for the corresponding grouping temperature. The color scale represents number of canonical keys.

\includepdf[
  pages=-,
  pagecommand={\thispagestyle{plain}},
  width=\textwidth
]{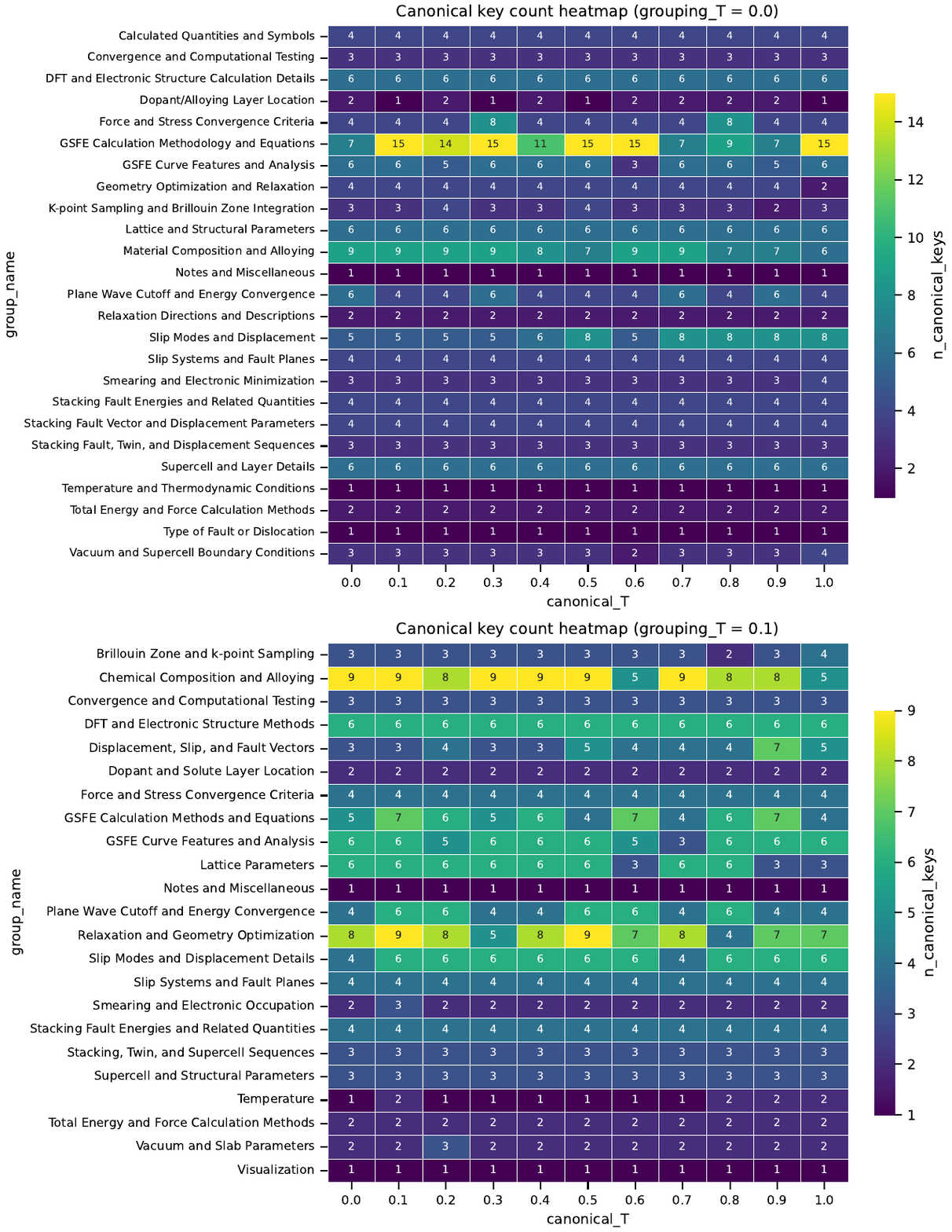}

\begin{figure}[p]
  \centering
  \label{fig:canonical_key_heatmaps}
\end{figure}
\end{document}